\documentclass[a4paper, 10pt]{article}
\usepackage{amsmath,amssymb,bm}
\usepackage[dvipdfmx]{graphicx}
\title{Mean Field Analysis of Stochastic Neural Network Models \\with Synaptic Depression}
\author{Yasuhiko Igarashi$^{1\,\dagger}$, Masafumi Oizumi$^{1,2\,\ddagger}$, Masato Okada$^{1,3\,\dagger\dagger}$
}
\begin{document}
\maketitle
\section*{Abstract}
We investigated the effects of synaptic depression on the macroscopic behavior of stochastic neural networks. Dynamical mean field equations were derived for such networks by taking the average of two stochastic variables: a firing state varialbe and a synaptic variable. In these equations, their average product is decoupled as the product of averaged them because the two stochastic variables are independent. We proved the independence of these two stochastic variables assuming that the synaptic weight $J_{ij}$ is of the order of $1/N$ with respect to the number of neurons $N$. 
Using these equations, we derived macroscopic steady state equations for a network with uniform connections and a ring attractor network with Mexican hat type connectivity and investigated the stability of the steady state solutions. An oscillatory uniform state was observed in the network with uniform connections due to a Hopf instability. With the ring network, high-frequency perturbations were shown not to affect system stability. Two mechanisms destabilize the inhomogeneous steady state, leading two oscillatory states. A Turing instability leads to a rotating bump state, while a Hopf instability leads to an oscillatory bump state, which was previous unreported. Various oscillatory states take place in a network with synaptic depression depending on the strength of the interneuron connections.
\footnote{$^1$ {\it Graduate School of Frontier Sciences, The University of Tokyo, Kashiwa, Chiba 277-8561}\\
$^2$ {\it Research Fellow of the Japan Society for the Promotion of Science}\\
$^3$ {\it Brain Science Institute, RIKEN, Wako, Saitama, 351-0198}\\
$\dagger$ E-mail address: igayasu1219@mns.k.u-tokyo.ac.jp\\
$\ddagger$ E-mail address: oizumi@mns.k.u-tokyo.ac.jp\\
$\dagger\dagger$E-mail address: okada@k.u-tokyo.ac.jp\\
}
\clearpage
\section{Introduction}

Neurophysiological experiments have shown that high-frequency inputs reduce the efficacy of signal transmission due to the exhaustion of neurotransmitters\cite{Thomson}. This phenomenon, called "synaptic depression", provides a dynamic gain control by reducing presynaptic inputs\cite{Abbott1997,Tsodyks,Rothman2009}. The synaptic depression not only affects the activity of a single neuron but also the overall activity in neural networks\cite{Abbott2}. To explore the effects of synaptic depression on the macroscopic behavior of stochastic neural networks, we analyzed a system based on a stochastic binary neural network model with synaptic depression by using mean-field theory. Preliminary results for the present work have been published elsewhere\cite{igarashi2009}.

To observe the macroscopic behavior of the network, we reduced the stochastic neural network model with synaptic depression to microscopic dynamical mean field equations by taking the average of two stochastic variables, a firing state variable and a synaptic variable, over different realizations of stochastic spikes. If the two stochastic variables are independent, their average product can be decoupled as the product of their average. This decoupling enables a stochastic network model with synaptic depression to be reduced the closed form equations of their average. We demonstrated such independence for a non-frustrated system and derived microscopic dynamical mean field equations for a stochastic binary neural network model with synaptic depression. The derived equations coincide with those of an analog neural network with synaptic depression\cite{Abbott1997,Tsodyks,Tsodyks1998}. 

Using these microscopic equations, we derived macroscopic steady state equations and analyzed the stability of macroscopic steady state solutions for two types of neural network : one with uniform connections and one with Mexican-hat type connections\cite{York,Kilpatrick2009_JCNS,Kilpatrick2009_PhysicaD}. A network with uniform connections is the simplest type of network, for which the effect of synaptic depression has been studied\cite{Tsodyks1998}. A ring neural network with Mexican-hat type connections has with nonuniform connectivity and has been used as a model of the primary visual cortex and prefrontal cortex\cite{BenYishai,Compte}. %
Although several researchers have reported that synaptic depression in the ring network leads to an oscillatory state, which is called a "rotating bump (RB)" state or a "traveling wave" state, little is known about the cause of the oscillation\cite{York,Kilpatrick2009_JCNS,Kilpatrick2009_PhysicaD}. 

We first discuss a network with uniform connections. Due to its simpleness, we can reduce the network to a model with only two components in which an oscillatory instability (a Hopf instability) can take place. This reduction enables us to easily understand why synaptic depression causes oscillatory instability. We first show that an oscillatory uniform (OU) state occurs due to the presence of synaptic depression. Although it has been reported that synaptic depression leads to an oscillatory state in a network with non-uniform connections\cite{York,Kilpatrick2009_JCNS,Kilpatrick2009_PhysicaD}, an oscillatory state has not been reported for a network with uniform connections. 

Next, we show that, in a ring network with Mexican-hat type connections, the synaptic depression leads to three oscillatory states: the OU state, the RB state, and an oscillatory bump (OB) state, which was previously unreported. 
To investigate the mechanisms of steady state destabilization that lead to these oscillatory states, we analyzed the stability of the steady state solutions when there are frequency perturbations. In \S\ref{sec:Network with MH}, we show that high-frequency perturbations do not affect the system stability and that two mechanisms destabilize an inhomogeneous steady state, leading to the two oscillatory states, i.e., the RB and the OB states. Analytical and numerical investigation of the ring network with synaptic depression revealed a phase diagram in which a wealth of states and multistable regimes can be discerned.

\section{Model}
\begin{figure}[!h]
\centering
\includegraphics[width=3in,clip]{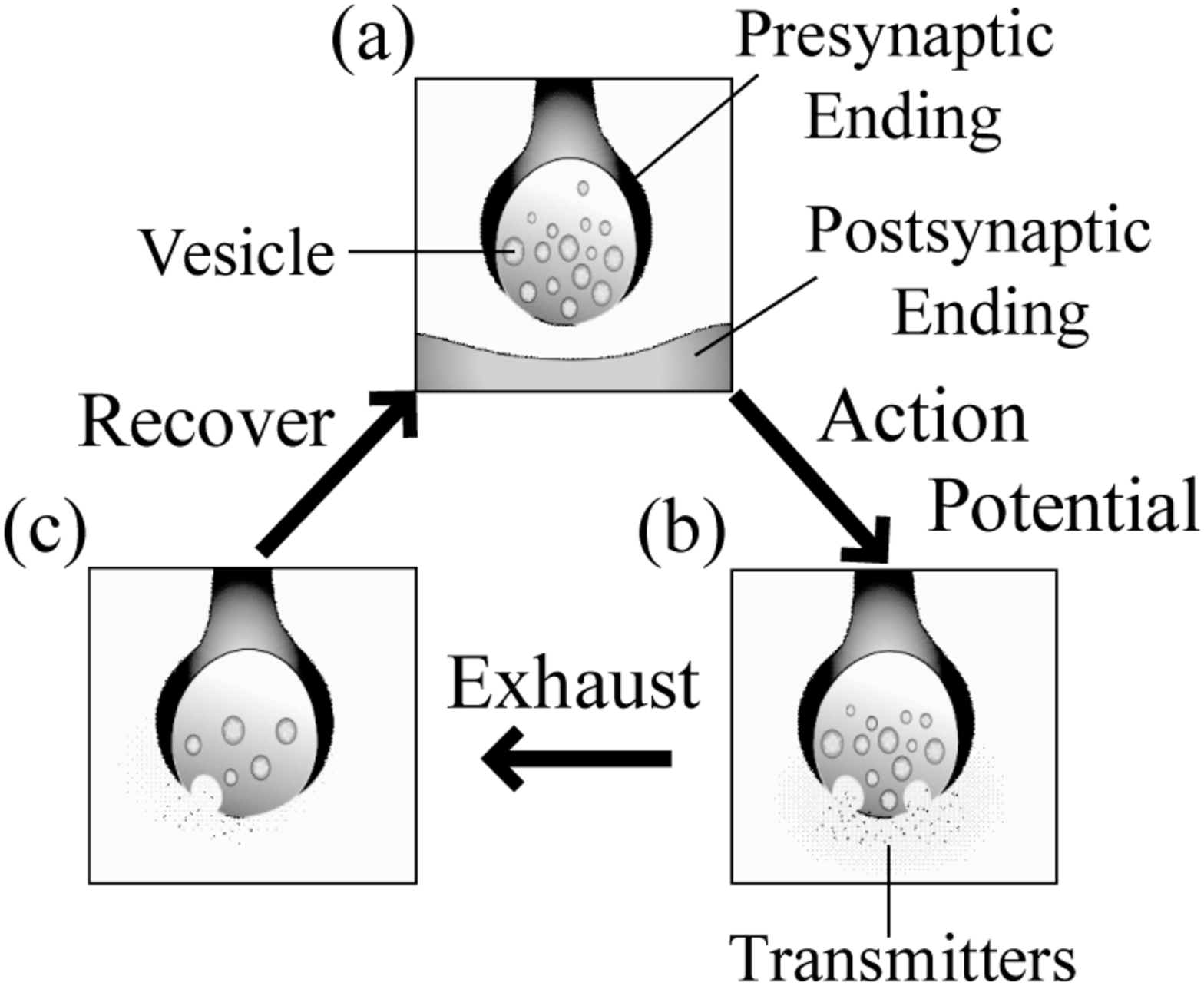}
\caption{Schematic of phenomenological model of synaptic depression}
\label{Fig:SD_BioMech}
\end{figure}

We used a recurrent neural network with $N$ neurons. The state of the $i$-th neuron at time $t$ is represented by $s_i(t)$. The state takes either a resting state $s_i(t)=0$ or a firing state $s_i(t)=1$. Each neuron follows a probabilistic dynamic:
\begin{eqnarray}
\mathrm{Prob}[s_i(t+1)=1]&\equiv& g_{\beta}(h_i(t)),\label{Prob_s(t+1)}\\
g_{\beta}(h_i(t))&=&\frac{1}{2}\bigl(1+\tanh(\beta h_i(t)\bigr)\label{g_beta_def}, 
\end{eqnarray}
where $h_i$($=\sum^{N}_{j\neq i}J_{ij}(2x_j(t)s_j(t)-1)$) represents the total synaptic current arriving at neuron $i$, and $1/\beta(=T)$ is the level of noise due to the stochastic synaptic activity. At each time step, all neurons are updated in parallel. $J_{ij}$ is a fixed synaptic weight from the $j$-th to the $i$-th neuron. $x_j(t)$($0<x_j(t)\leq1$) denotes the efficacy of signal transmission at the $j$-th neuron, which dynamically changes due to synaptic depression. $x_j(t)$ is determined by the corresponding neuron state and itself at preceding time $t-1$: 
\begin{align}
x_j(t)&=x_j(t-1)+\frac{1-x_j(t-1)}{\tau}-U_{\mathrm{se}}x_j(t-1)s_j(t-1).
\label{SD_model}
\end{align}
In this model, synaptic connection $J_{ij}(t)$($=J_{ij}x_j(t)$) dynamically changes with the efficacy of signal transmission $x_j(t)$. 
The phenomenological model of synaptic depression described by eq. (\ref{SD_model}) was proposed by several researchers\cite{Abbott1997,Tsodyks}. A schematic of this model is shown in Fig. \ref{Fig:SD_BioMech}. Information from one neuron (presynaptic neuron) flows to another neuron (postsynaptic neuron) across a synapse. A synapse is a small gap separating neurons and consists of a presynaptic ending that contains neurotransmitters stored in synaptic vesicles, a postsynaptic ending containing receptor sites for neurotransmitters, and a synaptic cleft, or space, between the presynaptic and postsynaptic endings (Fig. \ref{Fig:SD_BioMech}(a)). 
An action potential cannot cross the synaptic cleft between neurons. Instead the nerve impulse is carried by chemicals called neurotransmitters (Fig. \ref{Fig:SD_BioMech}(b)). The cell receiving the nerve impulse (the postsynaptic neuron) has chemical-gated ion channels, called neuroreceptors, in its membrane. The presynaptic neurons exhaust neurotransmitters when they transmit signals. The efficacy of signal transmission at presynaptic neuron $j$ at time $t$ decreases by a certain fraction, $U_{\mathrm{se}}x_j(t-1)$($0<U_{\mathrm{se}}\le1$), after the firing of the presynaptic neuron, $s_j(t-1)=1$ (Fig. \ref{Fig:SD_BioMech}(c)), and recovers with time constant $\tau$($\tau\ge1$), as shown in eq. (\ref{SD_model}).

\section{Mean field theory}
We propose a dynamical mean field theory for a stochastic binary neural network model with synaptic depression assuming that synaptic weight $J_{ij}$ is of the order of $1/N$ with respect to the number of neurons, $N$. 
\subsection{Microscopic mean field theory}
 We derived microscopic dynamical mean field equations by first taking the noise average of the firing state variable at time $t$: 
\begin{align}
\langle s_{i}(t+1)\rangle&=g_\beta(\langle h_i(t)\rangle),\\
\langle h_i(t)\rangle&=\sum_{j\neq i}^N J_{ij}(2\langle x_j(t)s_j(t)\rangle-1).\label{meanf_s2}
\end{align}
Similarly, we consider the noise average of eq. (\ref{SD_model}) for the synaptic variable:  
\begin{eqnarray}
\langle x_j(t+1)\rangle=\langle x_j(t)\rangle+\frac{1-\langle x_j(t)\rangle}{\tau}-U_{\mathrm{se}}\langle x_j(t)s_j(t)\rangle.
\label{meanf_x}
\end{eqnarray} 
Evidently, $x_i(t+\tau)$ and $s_i(t)$ are correlated when $\tau>0$. However, equal time correlations between $s_i(t)$ and $x_i(t)$ are of the order of $1/N$; that is, $x_i(t)$ and $s_i(t)$ become "independent" when $N\to\infty$, as we will show below.

Here, we define $\delta x_i(t)=x_i(t)-\langle x_i(t)\rangle$ and $\delta s_i(t)=s_i(t)-\langle s_i(t)\rangle$. Substituting eq. (\ref{Prob_s(t+1)}) for $\langle\delta x_i(t)\delta s_i(t)\rangle$, we obtain
\begin{align}
\langle\delta x_i(t)\delta s_i(t)\rangle=\langle\delta x_i(t)g_{\beta}\left(h_i(t-1)\right)\rangle.
\label{independent1}
\end{align}
Taylor expansion gives 
\begin{eqnarray}
g_{\beta}\left(h_i(t-1)\right)=g_{\beta}\left(\langle h_i(t-1)\rangle\right)
+g'_{\beta}\left(\left\langle h_i(t-1)\right\rangle\right)\delta\left(h_i(t-1)\right)+\dots.
\label{Tayor_approx}
\end{eqnarray}
Neglecting the higher order terms yields
\begin{eqnarray}
\langle\delta x_i(t)\delta s_i(t)\rangle&=&g'_{\beta}\left(\left\langle h_i(t-1)\right\rangle\right)\langle\delta h_i(t-1)\delta x_i(t)\rangle\\
&=&g'_{\beta}\left(\left\langle h_i(t-1)\right\rangle\right)
\sum_{j\neq i}^NJ_{ij}\langle\delta(x_j(t-1)s_j(t-1))\delta x_i(t)\rangle.
\label{independent2}
\end{eqnarray}
We evaluate the order of the right hand side of eq. (\ref{independent2}): 
\begin{multline}
\langle\delta(x_j(t-1)s_j(t-1))\delta x_i(t)\rangle=\langle x_j(t-1)\rangle
\langle\delta s_j(t-1)\delta x_i(t)\rangle\\
+\langle s_j(t-1)\rangle\langle\delta x_j(t-1)\delta x_i(t)\rangle+\langle\delta x_j(t-1)\delta s_j(t-1)\delta x_i(t)\rangle.
\label{sxj_xi_correlation}
\end{multline}
$\langle\delta(x_j(t-1)s_j(t-1))\delta x_i(t)\rangle$ is of the order of $1/N$, because time-delayed cross-correlation is of the order of $1/N$\cite{Ginzburg1994}. Since we have assumed $J_{ij}\sim O(1/N)$, we obtain 
\begin{eqnarray}
\sum_{j\neq i}^NJ_{ij}\langle\delta(x_j(t-1)s_j(t-1))\delta x_i(t)\rangle\sim O(1/N)
\end{eqnarray} 
and the equal-time correlations between $s_i(t)$ and $x_i(t)$, $\langle\delta x_i(t) \delta s_i(t) \rangle$, disappear in the limit of large networks, $N\to\infty$:
\begin{align}
\bigl\langle x_i(t)s_i(t)\bigr\rangle=\langle x_i(t)\rangle\langle s_i(t)\rangle.\label{independent}
\end{align}

Taking advantage of the independence between $x_i(t)$ and $s_i(t)$, we obtained the dynamical mean field equations for $m_i(t)$ and $X_i(t)$:
\begin{align}
m_i(t+1)&=g_{\beta}\left(\sum_{j\neq i}^N J_{ij}\left(2m_{j}(t)X_{j}(t)-1\right)\right),
\label{meanf_m}\\
X_i(t+1)&=X_i(t)+\frac{1-X_i(t)}{\tau}-U_{\mathrm{se}}X_i(t)m_i(t), 
\label{meanf_X}
\end{align}
where $m_i(t)\equiv\langle s_i(t)\rangle$ and $X_i(t)\equiv\langle x_i(t)\rangle$. 
These equations for the stochastic neural network model coincide with those for an analog neural network with synaptic depression\cite{Tsodyks1998}. The steady state equation for noise average $X_j=X_j(\infty)$ is 
\begin{eqnarray}
X_j=\frac{1}{1+\gamma m_j}, \quad\gamma=\tau U_{\mathrm{se}},
\end{eqnarray}
 which is a finite temperature version obtained in the $T=0$ case\cite{Matsumoto}. Finally, we obtain the microscopic steady state equation for $m_i$($=m_i(\infty$) for a network with synaptic depression:  
\begin{eqnarray}
m_i=g_{\beta}\left(\sum_{j\neq i}^N J_{ij}\left(\frac{2m_{j}}{1+\gamma m_j}-1\right)\right).
\label{self_eq}
\end{eqnarray}
According to eq. (\ref{self_eq}), steady state depends on $\gamma$($=\tau U_{\mathrm{se}}$)\cite{Tsodyks1998}.

\subsection{Stability analysis}
 \label{subsec:Stab Analysis}
To examine the stability of the steady state obtained with eq. (\ref{self_eq}), we consider small deviations around a fixed point\cite{Tsodyks1998,York,Hansel1998}: 
\begin{eqnarray}
m_i(t)=m_i+\delta m_i(t),\quad
X_i(t)=X_i+\delta X_i(t),
\end{eqnarray}
where $X_i=\frac{1}{1+\gamma m_i}$.
We linearize eqs. (\ref{meanf_m}) and (\ref{meanf_X}) about the steady state solution. Neglecting the higher order terms, we obtain 
\begin{eqnarray}
\delta m_i(t+1)&=&
\sum_{j\neq i}\Bigl(\frac{\partial g_{\beta}\left(h_i\right)}{\partial m_j}\delta m_j(t)+\frac{\partial g_{\beta}\left(h_i\right)}{\partial X_j}\delta X_j(t)\Bigr)
\label{FD_linear_eq_m},\\
h_i&=&\sum_{j\neq i}J_{ij}\left(2m_jX_j-1\right).
\end{eqnarray}
Similarly, 
\begin{eqnarray}
\delta X_i(t+1)=-U_\mathrm{se}X_{i}\delta m_i(t)+\left(1-\frac{1}{\tau}-U_\mathrm{se}m_i\right)\delta X_i(t).
\label{FD_linear_eq_X}
\end{eqnarray}
Next we calculate the partial differential coefficients $\frac{\partial g_{\beta}\left(h_i\right)}{\partial m_j}$ and $\frac{\partial g_{\beta}\left(h_i\right)}{\partial X_j}$ of eq. (\ref{FD_linear_eq_m}) in more detail.
\begin{eqnarray}
\frac{\partial g_{\beta}(h_i)}{\partial m_j}&=&\frac{\partial g_{\beta}(h_i)}{\partial h_i}\frac{\partial h_i}{\partial m_j}\\
&=&\beta J_{ij}X_j\left(1-(\tanh\beta(h_i))^2\right)
\label{partial_dg_dm}
\end{eqnarray}
Substituting the steady state equation. (\ref{self_eq}) into eq. (\ref{partial_dg_dm}), we obtain
\begin{eqnarray}
\frac{\partial g_{\beta}(h_i)}{\partial m_j}&=&4\beta J_{ij}X_j\left(m_i-m_i^2\right).
\label{partial_m_j}
\end{eqnarray}
Similarly, we have
\begin{eqnarray}
\frac{\partial g_{\beta}(h_i)}{\partial X_j}&=&4\beta J_{ij}m_j\left(m_i-m_i^2\right).\label{partial_X_j}
\end{eqnarray}
Substituting eqs. (\ref{partial_m_j}) and (\ref{partial_X_j}) into eq. (\ref{FD_linear_eq_m}) yields
\begin{eqnarray}
\delta m_i(t+1)&=&
\sum_{j\neq i}4\beta J_{ij}
\left(m_i-m_i^2\right)
\left(X_j\delta m_j(t)+m_j\delta X_j(t)\right).
\label{FD_linear_eq_m2}
\end{eqnarray}

From the relations for the coefficients of eqs. (\ref{FD_linear_eq_X}) and (\ref{FD_linear_eq_m2}), we obtain the Jacobian matrix for the system\cite{York}. The Jacobian matrix $K$ has a size of $2N\times2N$ with matrix elements as follows.
\begin{eqnarray}
K^{ij}\equiv
\begin{pmatrix} 
K_{{m}{m}}^{ij}&K_{{m}{X}}^{ij}\\
K_{{X}{m}}^{ij}&K_{{X}{X}}^{ij}
\end{pmatrix},\quad
\begin{pmatrix}
\delta m_i(t+1)\\
\delta X_i(t+1)
\end{pmatrix}
=K^{ij}
\begin{pmatrix}
\delta m_{j}(t)\\
\delta X_{j}(t)
\end{pmatrix},
\label{K_ij}
\end{eqnarray}
\begin{eqnarray}
K_{mm}^{ij}=4\beta J_{ij}(m_i-m_i^2)X_{j},\quad K_{mX}^{ij}=4\beta J_{ij}(m_i-m_i^2)m_{j},
\end{eqnarray}
\begin{eqnarray}
K_{Xm}^{ij}=-\delta_{ij}U_{\mathrm{se}}X_j,\quad K_{XX}^{ij}=\delta_{ij}\left(1-\frac{1}{\tau}\right)-U_{\mathrm{se}}m_j,
\end{eqnarray}
where $1\le i,j \le N$ and $\delta_{ij}$ is the Kronecker delta. If the Jacobian matrix has eigenvalues of $1$ or less, the steady state solution is stable. 

\section{Network with uniform connections}
\label{sec:Ferro Network}
In a network with uniform connections, 
\begin{eqnarray}
J_{ij}=J_0/N.
\label{J_ij=J_0/N}
\end{eqnarray}
\subsection{Macroscopic steady state equations}
We derived macroscopic steady state equations for a network with homogeneous connectivity by using the microscopic mean field equations (\ref{meanf_m}) and (\ref{meanf_X}). Given the symmetry of the synaptic weights in eq. (\ref{J_ij=J_0/N}), we can set the noise average of each neuron, $m_i$, to $m_i=\hat{m}_0$, where $\hat{m}_0=\frac{1}{N}\sum_{i=1}^Nm_i $. Substituting this condition into eq. (\ref{self_eq}), we obtain a macroscopic steady state equation:
\begin{eqnarray}
\hat{m}_0&=&g_{\beta}\left(J_0\left(\frac{2\hat{m}_0}{1+\gamma \hat{m}_0}-1\right)\right)\\
	&=&\frac{1}{2}\left(1+\tanh \beta J_0\left(\frac{2\hat{m}_0}{1+\gamma \hat{m}_0}-1\right)\right).
\label{FD_self_eq}
\end{eqnarray}
Eq. (\ref{FD_self_eq}) gives the homogeneous steady state solution.

\subsection{Stability analysis}
\label{sec:Stab Analysis Ferro}
To examine the stability of the homogeneous steady state solution obtained using eq. (\ref{FD_self_eq}), namely $m_i=\hat{m}_0$ and $X_i=\hat{X}_0$, we consider small deviations around a fixed point\cite{Tsodyks1998,York,Hansel1998}: 
\begin{eqnarray}
m_i(t)=\hat{m}_0+\delta m_i(t),\quad
X_i(t)=\hat{X}_0+\delta X_i(t).
\label{FD_small_deviations}
\end{eqnarray}
Substituting  eqs. (\ref{J_ij=J_0/N}) and (\ref{FD_small_deviations}) into eqs. (\ref{FD_linear_eq_m}) and (\ref{FD_linear_eq_m2}), we obtain
\begin{eqnarray}
\delta m_i(t+1)&=&
4\beta J_0
\left(\hat{m}_0-\hat{m}_0^2\right)
\frac{1}{N}\sum_{j\neq i}^N
\left(\hat{X}_0\delta m_j(t)+\hat{m}_0\delta X_j(t)\right).
\label{FD_linear_eq_m3}
\end{eqnarray}
\begin{eqnarray}
\delta X_i(t+1)=-U_\mathrm{se}\hat{X}_{0}\delta m_i(t)+\left(1-\frac{1}{\tau}-U_\mathrm{se}\hat{m}_0\right)\delta X_i(t).
\label{FD_linear_eq_X3}
\end{eqnarray}
Since $J_{ij}$ consists of $0$-th the Fourier component of $J_0$, 
discrete Fourier transform analysis can be use to diagonalize Jacobian matrix $K$ (eq. (\ref{K_ij})). We therefore compute the Fourier series for eqs. (\ref{FD_linear_eq_m3}) and (\ref{FD_linear_eq_X3}). 
The Fourier coefficients of $\delta m_i(t)$ and $\delta X_i(t)$ are given by
\begin{eqnarray}
\delta\hat{m}_k(t)=\frac{1}{N}\sum_{i=1}^{N}\delta m_i(t)e^{\frac{-2\pi\mathrm{i}(ik)}{N}},\quad \delta\hat{X}_k(t)=\frac{1}{N}\sum_{i=1}^{N}\delta X_i(t)e^{\frac{-2\pi\mathrm{i}(ik)}{N}}, 
\end{eqnarray}
where $\mathrm{i}$ is the standard imaginary unit with the property $\mathrm{i}^2=-1$. We then write $\delta m_i(t)$ and $\delta X_i(t)$ in Fourier series form: 
\begin{align}
\delta m_i(t)=\sum_{k=-\frac{N}{2}}^{\frac{N}{2}-1} \delta \hat{m}_k(t)e^{\frac{2\pi\mathrm{i}(ik)}{N}},\quad \delta X_i(t)=\sum_{k=-\frac{N}{2}}^{\frac{N}{2}-1} \delta \hat{X}_k(t)e^{\frac{2\pi\mathrm{i}(ik)}{N}}.
\label{m_j(t)=fourier}
\end{align}
Substituting eq. (\ref{m_j(t)=fourier}) for eqs. (\ref{FD_linear_eq_m3}) and (\ref{FD_linear_eq_X3}), we obtain
\begin{multline}
\sum_{k=-\frac{N}{2}}^{\frac{N}{2}-1} \delta \hat{m}_k(t+1)e^{\frac{2\pi\mathrm{i}(ik)}{N}}=
4\beta J_0
\left(\hat{m}_{0}-\hat{q}_{0}\right)
\sum_{l=-\frac{N}{2}}^{\frac{N}{2}-1}\left(\hat{X}_{0}\delta\hat{m}_l(t)+\hat{m}_{0}\delta \hat{X}_l(t)\right)
\frac{1}{N}\sum_{j\neq i}^Ne^{\frac{2\pi\mathrm{i}\left(jl\right)}{N}},
\label{FD_linear_eq_m_F3}
\end{multline}

\begin{eqnarray}
=
4\beta J_0
\left(\hat{m}_{0}-\hat{q}_{0}\right)
\left(\hat{X}_{0}\delta\hat{m}_0(t)+\hat{m}_{0}\delta \hat{X}_0(t)\right),
\label{FD_linear_eq_m_F4}
\end{eqnarray}

\begin{multline}
\sum^{\frac{N}{2}-1}_{k=-\frac{N}{2}} \delta \hat{X}_k(t+1)e^{\frac{2\pi\mathrm{i}(ik)}{N}}=
\left(1-\frac{1}{\tau}\right)
\sum^{\frac{N}{2}-1}_{l=-\frac{N}{2}}
\delta\hat{X}_l(t)e^{\frac{2\pi\mathrm{i}(il)}{N}}\\
-U_{se}\sum^{\frac{N}{2}-1}_{l,l'=-\frac{N}{2}}\left(\delta \hat{m}_{l}(t) \hat{X}_{l'}(t)+\hat{m}_{l}(t) \delta \hat{X}_{l'}(t)\right)e^{\frac{2\pi\mathrm{i}\left(\left(l+l'\right)i\right)}{N}},
\label{FD_linear_eq_X_F}
\end{multline}
where $\hat{q}_k=\frac{1}{N}\sum_{i=1}^{N}(m_i)^2e^{\frac{-2\pi\mathrm{i}(ik)}{N}}$. We use the following equation in the limit of $N\to\infty$ to integrate the right side of eq. (\ref{FD_linear_eq_m_F3}) with respect to $j$: 
\begin{eqnarray}
\frac{1}{N}\sum_{j\neq i}^Ne^{\frac{2\pi\mathrm{i}\left(jl\right)}{N}}
=\left\{ \begin{array}{ll}
1 & (l=0)  \\
0 & (l\neq0)
\end{array}\right.
\end{eqnarray}

Since Fourier components are orthonormal, we can equate the coefficients of the Fourier components on the left and right sides. From the relations for the coefficients of eqs. (\ref{FD_linear_eq_m_F4}) and (\ref{FD_linear_eq_X_F}), we obtain the Jacobian matrix for the system in Fourier space, $H^{kk}$: 
\begin{description}
\item[$k=0$]
\begin{eqnarray}
\begin{pmatrix}
\delta {m}_0(t+1)\\
\delta {X}_0(t+1)
\end{pmatrix}
&=&H^{00}
\begin{pmatrix}
\delta \hat{m}_0(t)\\
\delta \hat{X}_0(t)
\end{pmatrix}
\end{eqnarray}
\begin{eqnarray}
&=&
\begin{pmatrix} 
 4\beta J_0\left(\hat{m}_0-\hat{q}_0\right)\hat{X}_0 &  4\beta J_0\left(\hat{m}_0-\hat{q}_0\right)\hat{m}_0\\
-U_{\mathrm{se}}\hat{X}_0  & 1-\frac{1}{\tau}-U_{\mathrm{se}}\hat{m}_0
\end{pmatrix}
\begin{pmatrix}
\delta \hat{m}_0(t)\\
\delta \hat{X}_0(t)
\end{pmatrix},
\label{FD_H00}
\end{eqnarray}
\item[$\mid k\mid\ge1$]
\begin{eqnarray}
\begin{pmatrix}
\delta \hat{m}_k(t+1)\\
\delta \hat{X}_k(t+1)
\end{pmatrix}
&=&H^{kk}
\begin{pmatrix}
\delta \hat{m}_k(t)\\
\delta \hat{X}_k(t)
\end{pmatrix}
\\
&=&
\begin{pmatrix} 
 0 & 0\\
-U_{\mathrm{se}}\hat{X}_0 & 1-\frac{1}{\tau}-U_{\mathrm{se}}\hat{m}_0
\end{pmatrix}
\begin{pmatrix}
\delta \hat{m}_k(t)\\
\delta \hat{X}_k(t)
\end{pmatrix}.
\label{FD_Hkk}
\end{eqnarray}
\end{description}

This form makes it easy to analyze the stability of a steady state since, for any $k$, the time evolution of each equation pair ($\delta\hat{m}_{k}(t)$ and $\delta\hat{X}_k(t)$) decouples from all other equation pairs. Eigenvalue $\lambda_k$ of $H^{kk}$ determines whether a steady state solution is stable or not. The solution is stable if it is $1$ or less. 
The values for the eigenvalues, $\lambda_k(\neq0)$, are given by 
$\lambda_k=0$ and $\left(1-\frac{1}{\tau}-U_{\mathrm{se}}\hat{m}_0\right)$. 
Since $1\le\tau$, $0\le U_{\mathrm{se}}\le1$, $0\le\hat{m}_0\le1$, $\mid\lambda_k\mid\le1$, and $\hat{m}_k$, $\hat{X}_k$ $\to0$ as $t\to\infty$. 
We consider the stability to homogeneous perturbations $\delta\hat{m}_0$ and $\delta\hat{X}_0$, and obtain eigenvalue $\lambda_0$ of $H^{00}$:
\begin{multline}
\lambda_0^{\pm}=\frac{1}{2}\Biggl[
4\beta J_0\left(\hat{m}_0-\hat{q}_0\right)+\left(1-\frac{1}{\tau}-U_{\mathrm{se}}\hat{m}_0\right)\\
\pm\left\{ 
\left(4\beta J_0\left(\hat{m}_0-\hat{q}_0\right)+\left(1-\frac{1}{\tau}-U_{\mathrm{se}}\hat{m}_0\right)\right)^2
-4\left(4\beta J_0\left(\hat{m}_0-\hat{q}_0\right)\left(1-\frac{1}{\tau}\right)\right)
\right\}^{\frac{1}{2}}
\Biggr].
\label{FD_lambda}
\end{multline}
We find that two types of linear instability of the stationary uniform state are possible\cite{Roxin2005}. For $\mid \lambda_0^{\pm}\mid>1$ and $\mathrm{Im}(\lambda_0^{\pm})=0$, homogeneous perturbations $\delta\hat{m}_0$ and $\delta\hat{X}_0$ cause firing-rate instability. If $\mid \lambda_0^{\pm}\mid$ is more than $1$ and $\mathrm{Im}(\lambda_0^{\pm})\neq0$, homogeneous perturbations $\delta\hat{m}_0$ and $\delta\hat{X}_0$ yield a Hopf instability. This results in growing oscillations and instability. A small-amplitude limit-cycle periodic solution exists near the steady state solution.

\subsection{Results}
\begin{figure}[!h]	
\centering
\includegraphics[width=4.5in,clip]{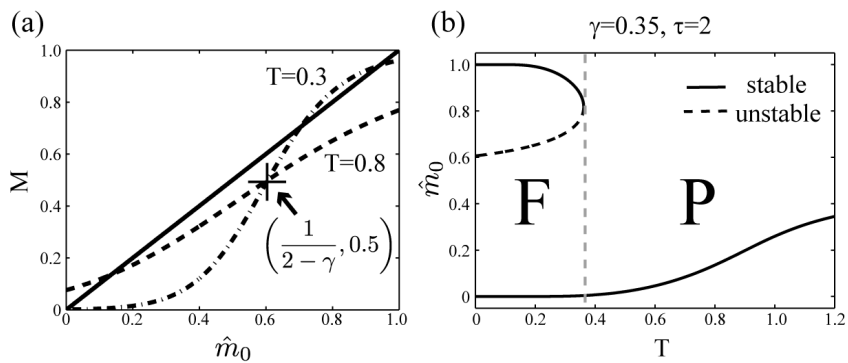}
\caption{(a) Graphical solutions to eq. (\ref{FD_self_eq}). (b) Stable and unstable solutions to eq. (\ref{FD_self_eq}) for $\gamma=0.35$, $\tau=2$.}
\label{fig:FD1}
\end{figure}

To evaluate the affect of synaptic depression on a network with uniform connections, we investigated the steady state solutions and their stability in the network. We obtained the steady state solutions to eq. (\ref{FD_self_eq}) by solving
\begin{align}
M&=\hat{m}_0,
\label{FD_self_eq_2}\\
M&=\frac{1}{2}\left(1+\tanh \beta J_0\left(\frac{2\hat{m}_0}{1+\gamma\hat{m}_0}-1\right)\right).
\label{FD_self_eq_3}
\end{align} 
Graphical solutions to eqs. (\ref{FD_self_eq_2}) and (\ref{FD_self_eq_3}) are shown in Fig. \ref{fig:FD1}(a) for $T=0.3$ (dot-dashed line) and $T=0.8$ (dashed line) when $\gamma=\tau U_{\mathrm{se}}=0.35$. Regardless of $\beta J_0$, the line for eq. (\ref{FD_self_eq_2}) passes through a point $(\frac{1}{2-\gamma}, 0.5)$ as shown in Fig. \ref{fig:FD1}(a). 
When the level of noise was low ($T=0.3$), eq. (\ref{FD_self_eq}) had three fixed points. Since two of them were attracting and the other was repelling, we found that the network with uniform connections had a bistable region at low and high $\hat{m}_0$ values in the case of low noise. We defined a ferromagnetic (F) state as bistability in $\hat{m}_0$\cite{Hamaguchi2008}. When the level of noise was high ($T=0.8$), the network with uniform connections had a monostable state for the $\hat{m}_0$ values. It is called a "paramagnetic (P)" state\cite{Hamaguchi2008}. When $\gamma>1$, $\frac{1}{2-\gamma}>1$ and the network had only a P state. 

The results of steady state solutions obtained using eq. (\ref{FD_self_eq}) and stability analysis for $\tau=2$ with a fixed degree of synaptic depression ($\gamma=0.35$) are shown in Fig. \ref{fig:FD1}(b). For $\gamma=0.35$ and $\tau=2$, the equilibrium point given by eq. (\ref{FD_self_eq}) was stable. The solid lines in Fig. \ref{fig:FD1}(b) denote stable solutions, and the dashed line denotes an unstable solution. A saddle node bifurcation occurred at $T_c=0.36$ since $\mid \lambda_0^{\pm}\mid >1$ and $\mathrm{Im} (\lambda_0^{\pm})=0$ (firing rate instability). Hence, the F state was stable for a low noise level ($T<T_c$), while the P state was stable for a high noise level ($T>T_c$). Using eq. (\ref{FD_lambda}), we found that all solutions with $\hat{m}_0\le0.5$ were stable. This means that synaptic depression stabilizes a low firing-rate state in a network with uniform connections.

\subsubsection{Hopf instability of homogeneous solution}
\label{subsubsec:Hopf instability of the homogeneous solution}
\begin{figure}[!h]	
\centering
\includegraphics[width=4.5in,clip]{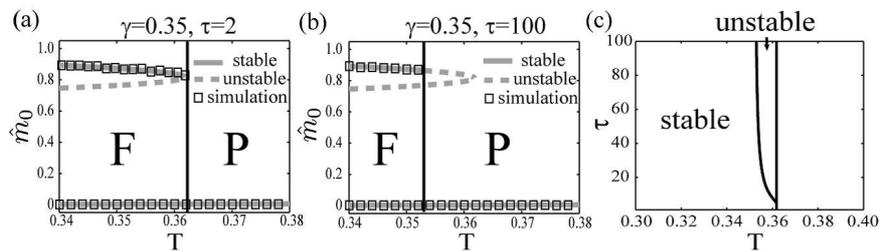}
\caption{Squares in figures (a) and (b) represent average $m$ numerically obtained by simulation with $N=10^3$. Solid lines denote stable solutions, and dashed lines denote unstable solutions. (c) Phase diagram for stability of $\hat{m}_0>0.5$ state.}
\label{fig:FD2}
\end{figure}

\begin{figure}[!h]	
\centering
\includegraphics[width=3in,clip]{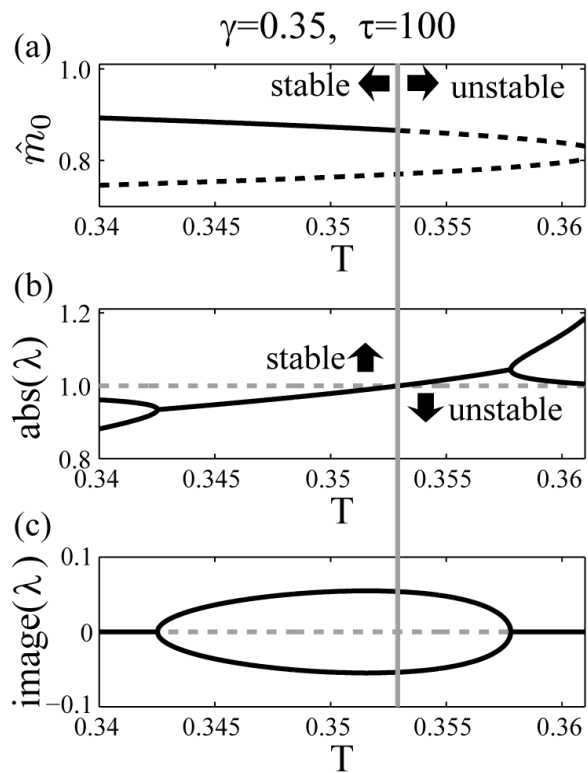}
\caption{Results of stability analysis for $\hat{m}_0>0.5$ and $0.34 \le T\le 0.362$. (a) Firing rate $\hat{m}_0$ state: solid line denotes stable solutions and dashed lines denote unstable solutions. (b)Absolute value of $\lambda_{0}^{\pm}$. (c)Image part of $\lambda_{0}^{\pm}$.}
\label{fig:FD_lambda_distrib}
\end{figure}

\begin{figure}[!h]
\centering
\includegraphics[width=4in,clip]{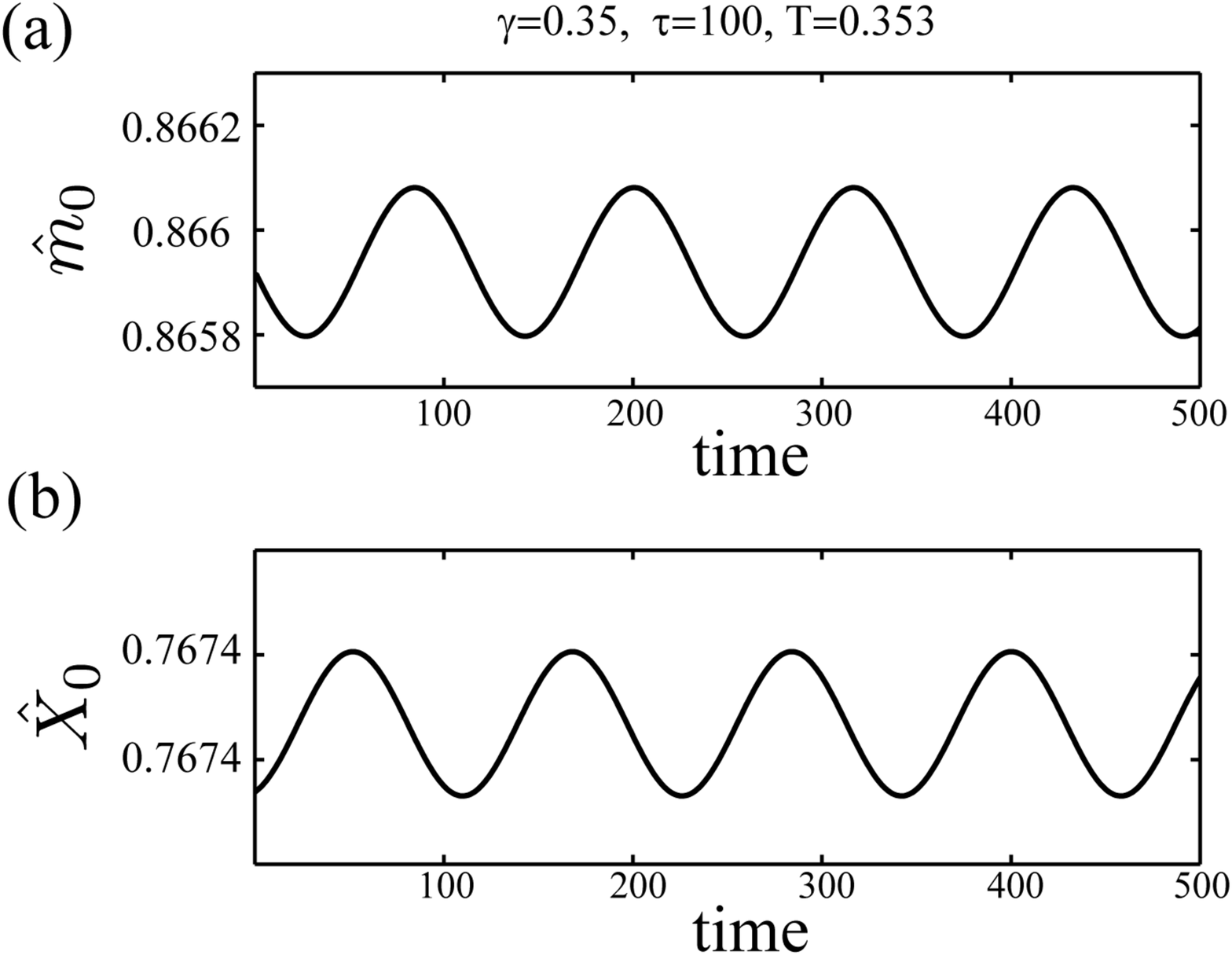}
\caption{(a) Firing rate $\hat{m}_0(t)$ and (b) average of synaptic efficacy $\hat{X}_0(t)$ corresponding to OU state.}
\label{fig:FD_trans_wave}
\end{figure}

We investigate the effect of the time constant $\tau$ on the stability of steady state solutions. We set $\gamma$ to $0.35$ and $\tau$ to $2$ and $100$. 
The steady state solutions obtained using eq. (\ref{FD_self_eq}) and stability analysis are shown in Fig. \ref{fig:FD2}(a) for $\tau=2$ and in Fig. \ref{fig:FD2}(b) for $\tau=100$. The simulation results are represented by the squares. The number of neurons was $10^3$, and the initial conditions for the firing state variables and synaptic variables were $s_i(0)=1$ and $x_i(0)=1$ for all $i$ and $s_i(0)=0$ and $x_i(0)=1$ for all $i$. 
Since $\gamma=0.35$ in both cases, the same steady state solutions are the same. We found that, near the transition point between the P and F states, a fixed point ($\hat{m}_0>0.5$) was stable for $\tau=2$ (Fig. \ref{fig:FD2}(a)), but unstable for $\tau=100$ (Fig. \ref{fig:FD2}(b)). Since $\mid \lambda_0^{\pm}\mid >1$ and $\mathrm{Im}( \lambda_0^{\pm})\neq0$ at $\hat{m}_0=0.865$ and $T=0.353$, as shown in Figs. \ref{fig:FD_lambda_distrib}(b) and (c), a Hopf bifurcation occurred, and there was an OU state near $T=0.353$. Fig. \ref{fig:FD_trans_wave} shows $\hat{m}_0(t)$($=\sum^N_i\hat{m}_i(t)$) and $\hat{X}_0(t)$($=\sum^N_i\hat{X}_i(t)$) for the OU state, respectively given by eqs. (\ref{meanf_m}) and (\ref{meanf_X}) ($N=10^3$). When $T$ was more than $0.353$, the OU state was unstable, and the macroscopic property of the network changed to the P state. These results are consistent with the simulation. 
The phase diagram of the stability for $\hat{m}_0>0.5$ shown in Fig. \ref{fig:FD2}(c) reveals that the state tended to be unstable when the time constant of the synaptic variable increased. The unstable region expanded until $\tau\approx50$, and then it settled in an approximately constant region. These results show that the stability of the solution strongly depends on the time constant, $\tau$.

Finally, we discuss the oscillatory instability in the neural network with uniform connections in the presence of synaptic depression. A neural network with synaptic depression can be considered an activator-inhibitor system in which firing rate $m$ is an activator and synaptic variable $X$ is an inhibitor. This is because inhibitor $X$ activated by $m$ and inhibits activator production since total synaptic input $h$ decreases as $X$ increases. In the activator-inhibitor system, an oscillatory instability (a Hopf instability) can take place\cite{Kuramoto2003,Murray2004}. We found that the oscillatory uniform state remained for a sufficiently long time constant in a neural network with uniform connections in the presence of synaptic depression. 
\section{Ring Attractor Network with Mexican-hat type connectivity}
\label{sec:Network with MH}
For a network with uniform connections, we found that the homogeneous state solution is unstable and that oscillatory uniform state remains for a sufficiently long time constant. 

Let us turn now to a ring attractor neural network with Mexican-hat type connectivity. In this network, $N$ neurons are located on a one-dimensional ring parameterized with $\theta$ $\in$ $[0, 2\pi)$. Neuron $i$ is labeled using angle $\theta_i$($=\frac{\pi i}{N}-\frac{\pi}{2}$). The synaptic weight is 
\begin{eqnarray}
J_{ij}=\frac{J_0}{N}+\frac{J_1}{N}\cos2(\theta_i-\theta_j),
\label{J_ij_MH} 
\end{eqnarray}
where $J_0$ is a uniform interaction, and $J_1$ is a lateral inhibition interaction \cite{BenYishai,Hamaguchi2006,Hamaguchi2008}.The model with $J_1=0$ reduces to a network with uniform connections. For $J_0=0$, the network is a balanced network with Mexican-hat type connectivity, like that studied by York {\it et al}\cite{York}. 

\subsection{Macroscopic steady state equation}
Using the microscopic mean field equations (\ref{meanf_m}) and (\ref{meanf_X}), we obtain macroscopic steady state equations for the ring network with synaptic depression.
Substituting eq. (\ref{J_ij_MH}) into the microscopic steady state equation (\ref{self_eq}), we obtain a microscopic equation: 
\begin{eqnarray}
m_i&=&g_{\beta}(h_i),\quad r_i=\frac{m_i}{1+\gamma m_i},\quad\\
h_i&=&J_0r_0+J_1\left(\frac{(r_1+r_{-1})}{2}\cos(2\theta_i)+\frac{(r_1-r_{-1})}{2i}\sin(2\theta_i)\right),
\end{eqnarray}
where
\begin{eqnarray}
r_0=\frac{1}{N}\sum^N_{i=1}2r_i-1,\quad 
r_1=\frac{1}{N}\sum^N_{i=1}(2r_i-1)e^{-2\mathrm{i}\theta_i},\quad 
r_{-1}=\frac{1}{N}\sum^N_{i=1}(2r_i-1)e^{2\mathrm{i}\theta_i}.
\end{eqnarray}
 The renormalized output of the $i$-th neuron due to the synaptic depression is $r_i=m_i/(1+\gamma m_i)$. We can obtain the following self-consistent equation for the macroscopic steady state in the limit $N\to\infty$,
\begin{eqnarray}
r_0=\frac{1}{\pi}\int_{-\frac{\pi}{2}}^{\frac{\pi}{2}}d\theta 2r(\theta)-1,\quad
r_1=\frac{1}{\pi}\int_{-\frac{\pi}{2}}^{\frac{\pi}{2}}d\theta 2r(\theta)e^{-2\mathrm{i}\theta},\quad
r_{-1}=\frac{1}{\pi}\int_{-\frac{\pi}{2}}^{\frac{\pi}{2}}d\theta 2r(\theta)e^{2\mathrm{i}\theta},
\end{eqnarray}
\begin{eqnarray}
r(\theta)=\frac{g_{\beta}\left(J_0r_0+J1\left(r_1e^{-2\mathrm{i}\theta}+r_{-1}e^{2\mathrm{i}\theta}\right)\right)}{1+\gamma g_{\beta}\left(J_0r_0+J1\left(r_1e^{-2\mathrm{i}\theta}+r_{-1}e^{2\mathrm{i}\theta}\right)\right)}.
\label{MD_selfeq_macro}
\end{eqnarray}
The order parameters given by eq. (\ref{MD_selfeq_macro}) are used to calculate the firing rate ($\hat{m}_0=\frac{1}{\pi}\int_{-\frac{\pi}{2}}^{\frac{\pi}{2}}d\theta m(\theta)$) and the $1$-st order Fourier component of $m(\theta)$ ($\hat{m}_1=\frac{1}{\pi}\int_{-\frac{\pi}{2}}^{\frac{\pi}{2}}d\theta m(\theta)e^{-2\mathrm{i}\theta}$). This component, $\hat{m}_1$, indicates the degree of activity localization.

There are two types of solutions to eq. (\ref{MD_selfeq_macro}). One is a homogeneous solution with $\hat{m}_1=0$, and the second one is a bump solution with $\hat{m}_1\neq0$, which is inhomogeneous.
\subsection{Stability analysis}
We investigated the stability of the steady state solution given by eq. (\ref{MD_selfeq_macro}) for the ring network as done for a network with uniform connections\cite{Tsodyks1998,York,Hansel1998}. To examine the stability of the steady state solution, ${m}_i$ and $X_i$, obtained using the eq. (\ref{MD_selfeq_macro}), we considered small deviations around a fixed point in eq. (\ref{MD_selfeq_macro}):
\begin{eqnarray}
m_i(t)=m_i+\delta m_i(t),\quad X_i(t)=X_i+\delta X_i(t).
\end{eqnarray}
Since $J_{ij}$ consists of the $0$-th Fourier component of $J_0$ and the $1$-st Fourier component of $J_1$ in the ring network, discrete Fourier transform analysis can be a used to diagonalize Jacobian matrix $K$ (eq. (\ref{K_ij})) for the ring network as well as for a network with uniform connections. We hence write $\delta m_i(t)$ and $\delta X_i(t)$ in Fourier series form in a way similar to that used in \S \ref{sec:Stab Analysis Ferro}. 
Substituting eqs. (\ref{m_j(t)=fourier}) and (\ref{J_ij_MH}) and $\theta_i=\frac{\pi i}{N}-\frac{\pi}{2}$ into eqs. (\ref{FD_linear_eq_X}) and (\ref{FD_linear_eq_m2}), we obtain
\begin{multline}
\sum_{k=-\frac{N}{2}}^{\frac{N}{2}-1} \delta \hat{m}_k(t+1)e^{2\mathrm{i}k\theta_i}=
\frac{1}{N}\sum_{j\neq i}^N\left(4\beta J_0+4\beta J_1\cos2(\theta_i-\theta_j)\right)\\
\times\sum_{k',l,l'=-\frac{N}{2}}^{\frac{N}{2}-1} 
\left(\hat{m}_{k'}-\hat{q}_{k'}\right)
\left(\hat{X}_{l'}\delta\hat{m}_l(t)+\hat{m}_{l'}\delta \hat{X}_l(t)\right)e^{2\mathrm{i}\left(k'\theta_i+\left(l+l'\right)\theta_j\right)}, 
\label{MD_linear_eq_m}
\end{multline}
\begin{multline}
\quad\quad\quad\quad\quad\quad\quad\quad\quad=
\sum^{\frac{N}{2}-1}_{k'=-\frac{N}{2}}\left(\hat{m}_{k'}-\hat{q}_{k'}\right)
\Biggl[
4\beta J_0e^{2\mathrm{i}k'\theta_i}\sum^{\frac{N}{2}-1}_{l=-\frac{N}{2}}\hat{m}_{l}(t)\hat{X}_{-l}(t)\\
+2\beta J_1e^{2\mathrm{i}\left(k'+1\right)\theta_i}\sum^{\frac{N}{2}-1}_{l=-\frac{N}{2}}\hat{m}_{l}(t)\hat{X}_{-(l-1)}(t)
+2\beta J_1e^{-2\mathrm{i}\left(k'-1\right)\theta_i}\sum^{\frac{N}{2}-1}_{l=-\frac{N}{2}}\hat{m}_{l}(t)\hat{X}_{-(l+1)}(t)
\Biggr],
\label{MD_linear_eq_m2}
\end{multline}
\begin{multline}
\sum^{\frac{N}{2}-1}_{k=-\frac{N}{2}} \delta \hat{X}_k(t+1)e^{2\mathrm{i}k\theta_i}=
\left(1-\frac{1}{\tau}\right)\sum^{\frac{N}{2}-1}_{l=-\frac{N}{2}}
\delta\hat{X}_l(t)e^{2\mathrm{i}l\theta_i}\\
-U_{se}\sum^{\frac{N}{2}-1}_{l,l'=-\frac{N}{2}}\left(\delta \hat{m}_{l}(t) \hat{X}_{l'}(t)+\hat{m}_{l}(t) \delta \hat{X}_{l'}(t)\right)e^{2\mathrm{i}\left(l+l'\right)\theta_i}, 
\label{MD_linear_eq_X}
\end{multline}
where we use the following equation in the limit of $N\to\infty$ to integrate the right side of eq. (\ref{MD_linear_eq_m}) with respect to $\theta_j$,
\begin{eqnarray}
\frac{1}{N}\sum_{j\neq i}^Ne^{2\mathrm{i}\left(l\theta_j\right)}
=\left\{ \begin{array}{ll}
1 & (l=0) \\
0 & (l\neq0)
\end{array}\right.
\end{eqnarray}

Since Fourier components are orthonormal, we can equate the coefficients of the Fourier components on the left and right sides. From the relations for the coefficients in eqs. (\ref{MD_linear_eq_X}) and (\ref{MD_linear_eq_m2}), we can obtain the Jacobian matrix for the system in Fourier space, $H$. The matrix has a size of $2N \times 2N$, with matrix elements
\begin{eqnarray}
H^{kl}\equiv
\begin{pmatrix} 
H_{\hat{m}\hat{m}}^{kl}&H_{\hat{m}\hat{X}}^{kl}\\
H_{\hat{X}\hat{m}}^{kl}&H_{\hat{X}\hat{X}}^{kl}
\end{pmatrix},\quad
\begin{pmatrix}
\delta \hat{m}_{k}(t+1)\\
\delta \hat{X}_{k}(t+1)
\end{pmatrix}
=H^{kl}
\begin{pmatrix}
\delta \hat{m}_{l}(t)\\
\delta \hat{X}_{l}(t)
\end{pmatrix},
\label{H}
\end{eqnarray}
\begin{eqnarray}
H_{\hat{m}\hat{m}}^{kl}=
4\beta J_0(\hat{m}_{k}-\hat{q}_{k})\hat{X}_{-l}
+2\beta J_1(\hat{m}_{k-1}-\hat{q}_{k-1})\hat{X}_{-(l-1)}
+2\beta J_1(\hat{m}_{k+1}-\hat{q}_{k+1})\hat{X}_{-(l+1)},
\end{eqnarray}
\begin{eqnarray}
H_{\hat{m}\hat{X}}^{kl}=
4\beta J_0(\hat{m}_{k}-\hat{q}_{k})\hat{m}_{-l}
+2\beta J_1(\hat{m}_{k-1}-\hat{q}_{k-1})\hat{m}_{-(l-1)}
+2\beta J_1(\hat{m}_{k+1}-\hat{q}_{k+1})\hat{m}_{-(l+1)},
\end{eqnarray}
\begin{eqnarray}
H_{\hat{X}\hat{m}}^{kl}=-U_{\mathrm{se}}\hat{X}_{k-l}, 
\end{eqnarray}
\begin{eqnarray}
H_{\hat{X}\hat{X}}^{kl}=\delta_{kl}\left(1-\frac{1}{\tau}\right)-U_{\mathrm{se}}\hat{m}_{k-l},
\end{eqnarray}
where $-\frac{N}{2}\le k, l\le \frac{N}{2}-1$, and $\delta_{kl}$ is the Kronecker delta. If the Jacobian matrix has eigenvalues of $1$ or less, the steady solution is stable. 

First, we consider the stability of homogeneous steady state ($\hat{m}_k=0$ and $\hat{X}_k=0$ ($k\neq0$)), which can be analytically analyzed as shown below\cite{York}. If the network has a homogeneous steady solution, $\hat{X}_k=0$, $\hat{m}_k=0$ ($k\neq0$), and we have
\begin{eqnarray}
H^{kl}=
\begin{pmatrix}
0&0\\
0&0
\end{pmatrix}\quad(k\neq l).
\label{H_kl_0}
\end{eqnarray}
This equation shows that the time evolution of each equation pair ($\delta\hat{m}_k(t)$ and $\delta\hat{X}_k(t)$) decouples from all other equation pairs. The Jacobian matrix for the ring network is therefore as easy to analyze as that for a network with uniform connections. The Jacobian matrix thus reduces to the following matrices:
\begin{description}
\item[$k=0$]
\begin{eqnarray}
H^{00}=
\begin{pmatrix}
4\beta J_0\left(\hat{m}_0-\hat{q}_0\right)\hat{X}_0 &  
4\beta J_0\left(\hat{m}_0-\hat{q}_0\right)\hat{m}_0\\
-U_{\mathrm{se}}\hat{X}_0  &  
1-\frac{1}{\tau}-U_{\mathrm{se}}\hat{m}_0
\end{pmatrix}
\label{H_00}
\end{eqnarray}
\item[$k=\pm1$]
\begin{eqnarray}
H^{11}&=&
\begin{pmatrix} 
 2\beta J_1\left(\hat{m}_0-\hat{q}_0\right)\hat{X}_0 &  
2\beta J_1\left(\hat{m}_0-\hat{q}_0\right)\hat{m}_0\\
-U_{\mathrm{se}}\hat{X}_0  &  
1-\frac{1}{\tau}-U_{\mathrm{se}}\hat{m}_0
\end{pmatrix}
\label{H_11}\\
H^{-1-1}&=&H^{11}
\end{eqnarray}
\item[$\mid k\mid >1$]
\begin{eqnarray}
H^{kk}=
\begin{pmatrix}  
0 &0\\
-U_{\mathrm{se}}\hat{X}_0& 1-\frac{1}{\tau}-U_{\mathrm{se}}\hat{m}_0
\end{pmatrix}
\end{eqnarray}
\end{description}
Since the eigenvalue $\lambda_k$ of $H^{kk} (\mid k\mid >1)$ is less than $1$,  $\delta\hat{m}_k$, $\delta\hat{X}_k$ $\to0$ as $t\to\infty$. We consider the stability under perturbations $\delta\hat{m}_0$, $\delta\hat{X}_0$, $\delta\hat{m}_1$, and $\delta\hat{X}_1$. The stability under perturbations $\delta\hat{m}_{-1}$ and $\delta\hat{X}_{-1}$ is identical to that under perturbations $\delta\hat{m}_1$ and $\delta\hat{X}_1$. The eigenvalue of $H^{00}$ is given by eq. (\ref{FD_lambda}). Next, we obtain eigenvalue $\lambda_1^{\pm}$ of $H^{11}$:
\begin{multline}
\lambda_1^{\pm}=\frac{1}{2}\Biggl[
2\beta J_1\left(\hat{m}_0-\hat{q}_0\right)+\left(1-\frac{1}{\tau}-U_{\mathrm{se}}\hat{m}_0\right)\\
\pm\left\{ 
\left(2\beta J_1\left(\hat{m}_0-\hat{q}_0\right)+\left(1-\frac{1}{\tau}-U_{\mathrm{se}}\hat{m}_0\right)\right)^2
-4\left(2\beta J_1\left(\hat{m}_0-\hat{q}_0\right)\left(1-\frac{1}{\tau}\right)\right)\right\}^{\frac{1}{2}}
\Biggr]
\label{homo_lambda1}
\end{multline}
The stability of the homogeneous steady state solution is determined by eqs. (\ref{FD_lambda}) and (\ref{homo_lambda1}). There are four types for linear instability of the homogeneous state: (1) firing-rate instability ($\mid \lambda_0^{\pm}\mid >1, \mathrm{Im}(\lambda_0^{\pm})=0$, $\mid \lambda_1^{\pm}\mid<1$), (2) Hopf instability ($\mid \lambda_0^{\pm}\mid >1, \mathrm{Im}(\lambda_0^{\pm})\neq0$, $\mid \lambda_1^{\pm}\mid<1$), (3) Turing instability ($\mid \lambda_0^{\pm}\mid<1$, $\mid \lambda_1^{\pm}\mid>1$, $\mathrm{Im}(\lambda_1^{\pm})=0$), and (4) Turing-Hopf instability ($\mid \lambda_0^{\pm}\mid<1$, $\mid \lambda_1^{\pm}\mid>1$, $\mathrm{Im}(\lambda_1^{\pm})\neq0$)\cite{Roxin2005}. If there is a Turing-Hopf instability, a spatially homogeneous steady state solution is unstable and, spatial periodic patterns evolve.

In contrast to the homogeneous steady state solutions, it is difficult to analyze the stability of inhomogeneous steady state solutions since the time evolution of each equation pair ($\hat{m}_{k}(t)$ and $\hat{X}_k(t)$) is coupled with other equation pairs. We hence have to take into account stability under frequency perturbations $\delta\hat{m}_k$ and $\delta\hat{X}_k$ ($-N/2\le k,l \le N/2-1$) since the highest Fourier component is $\hat{m}_{\pm N/2}$, $\hat{X}_{\pm N/2}$ in a network with $N$ by sampling theorem. Here we study how frequency perturbations $\delta\hat{m}_k$ and $\delta\hat{X}_k$ ($-N/2\le k,l \le N/2-1$) affect the stability of the inhomogeneous steady state solutions by considering the eigenvector of the Jacobian matrix (eq. (\ref{H})).

\subsection{Results}

\subsubsection{Oscillatory states}
\label{subsubsec:Oscillatory states}
\begin{figure}[!h]
\centering	
\includegraphics[width=4.5in,clip]{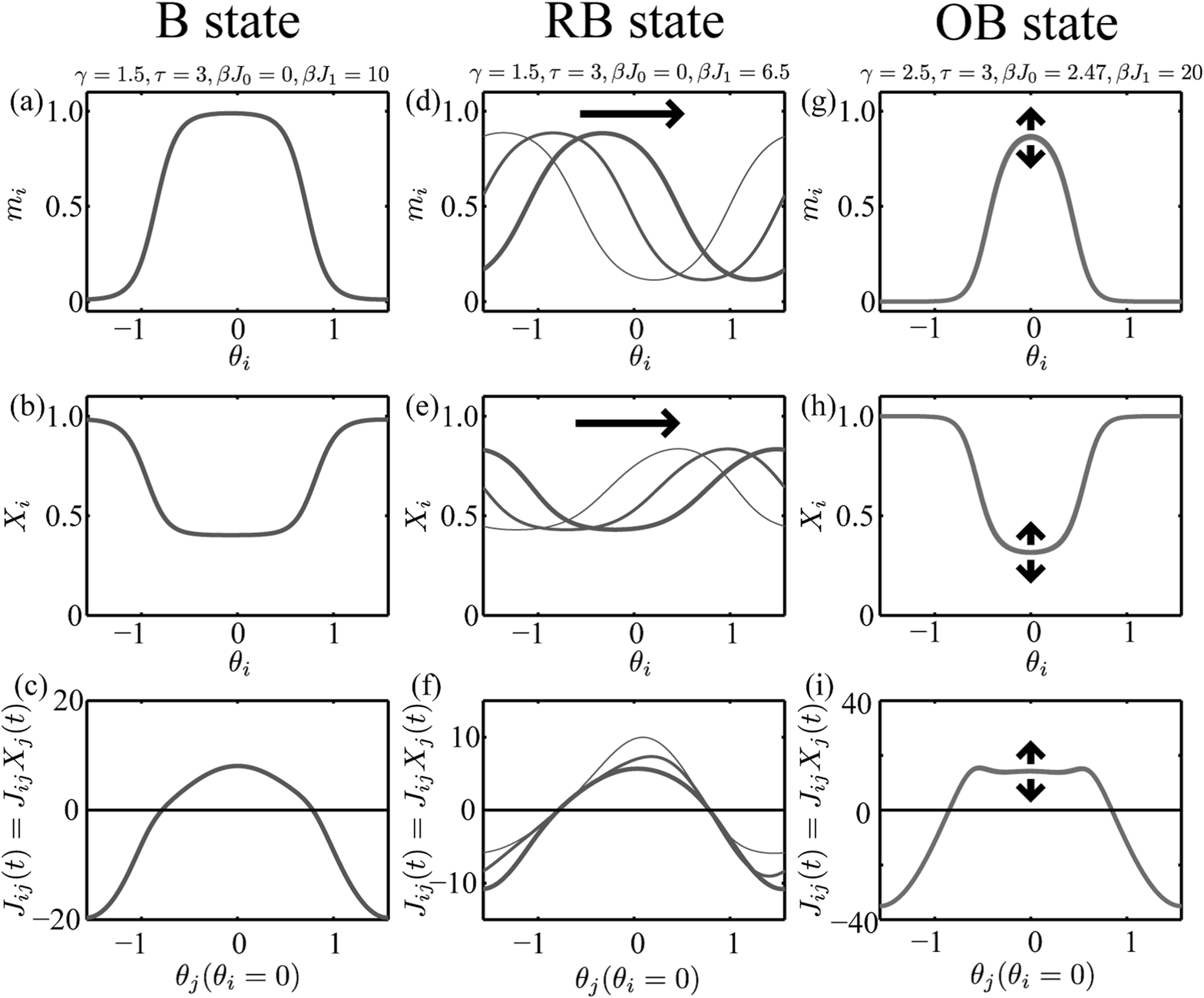}
\caption{Profiles for three inhomogeneous states (B, RB, OB). (a), (d), (g) Firing rate $m_i$. (b), (e), (h) Average of synaptic variable $X_i$. (c), (f), (i) Synaptic weight $J_{ij}(t)=J_{ij}X_{j}(t)$ between $i$-th neuron with preferred orientation $\theta_i=0$ and other neurons.}
\label{fig:MD_mj_Xj_Jij}
\end{figure}

Our investigation of the stability of the steady state solution given by eq. (\ref{MD_selfeq_macro}) for the ring network revealed six states in the network. Three are homogeneous and were also found in a network with uniform connections: a ferromagnetic (F) state, a paramagnetic (P) state, and an oscillatory uniform (OU) state. The OU state occurs in a way similar to that described in \S\ref{subsubsec:Hopf instability of the homogeneous solution} (Fig. \ref{fig:FD_lambda_distrib}). The other three states are inhomogeneous: a bump (B) state, a rotating bump (RB) state\cite{York,Kilpatrick2009_PhysicaD}, and an oscillatory bump (OB) state, as shown in Fig. \ref{fig:MD_mj_Xj_Jij}. The B state can be obtained using a self-consistent equation, eq. (\ref{MD_selfeq_macro}), while the other two cannot because they are dynamic states resulting from the destabilization of steady states. We thus obtained them by using the dynamical mean field equations, (\ref{meanf_m}) and (\ref{meanf_X}) with $N=10^4$. The firing rate $m_i$, average of the synaptic variable $X_i$, and synaptic weight between $i$-th neuron with the preferred orientation $\theta_i=0$ and the other neurons are shown in Fig. \ref{fig:MD_mj_Xj_Jij} for the three inhomogeneous states.

First, we discuss the behavior of the three inhomogeneous states in a ring network with synaptic depression. 
The B state is formed by a subset of the neurons firing in a self-reinforcing manner, causing localized activity (Fig. \ref{fig:MD_mj_Xj_Jij}(a)), similar to the B state in the network without synaptic depression\cite{BenYishai,Hamaguchi2006,York}. In the B state, the firing rates are high, while the averages of the synaptic variables are low (Figs. \ref{fig:MD_mj_Xj_Jij}(a) and (b)). 
The synaptic depression thus reduces the excitatory localized interaction and reduces the presynaptic inputs of the activated neuron, as shown in Fig. \ref{fig:MD_mj_Xj_Jij}(c).
In the RB state, a localized bump of activity propagates around the ring network, leaving a wake of replenishing synaptic resource, as shown in Figs. \ref{fig:MD_mj_Xj_Jij} (d) and (e)\cite{York}. In this example, the profile is moving to the right. As a result, the synaptic weights dynamically changed in the RB state (Fig. \ref{fig:MD_mj_Xj_Jij}(f)). 
In the OB state, which is first reported here, the bump state activity moved up and down around the neurons with the firing rates that were high, as shown in Figs. \ref{fig:MD_mj_Xj_Jij}(g) and (h). Unlike in the RB state, the moves are tiny in the OB state (Figs. \ref{fig:MD_mj_Xj_Jij}(g), (h) and (i)). The occurrence of the two oscillatory states from the B state implies that there are two mechanisms destabilizing the B state.

\begin{figure}[!h]
\centering	
\includegraphics[width=3in,clip]{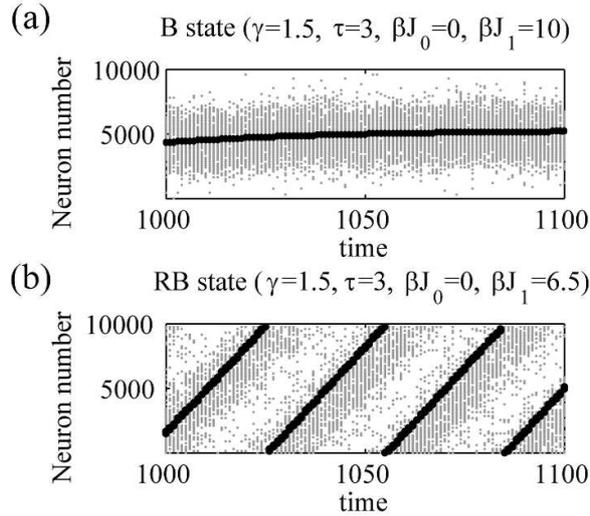}
\caption{Raster plots of neuron activity for numerical simulation ($N=10^4$). Solid lines denote temporal behaviors of bump position $\phi$. (a) B state. (b) RB state.}
\label{fig:raster}
\end{figure}

Next, we provide evidence that the simulation results coincide with the dynamic solution obtained using eqs (\ref{meanf_m}) and (\ref{meanf_X}) for the B state ($\beta J_0=0, \beta J_1=10$) and the RB state ($\beta J_0=0, \beta J_1=6.5$). Figs. \ref{fig:raster}(a) and (b) show raster plots of neuron activities obtained by numerical simulation with $N=10^4$. The solid lines represents the dynamic solution obtained using eqs. (\ref{meanf_m}) and (\ref{meanf_X}). In the B state, the localized position fluctuated and moved around the ring network since the B state is stable anywhere in the ring (Fig. \ref{fig:raster}(a))\cite{Hamaguchi2006}. In the RB state, the bumps propagated stably, i.e., there was a traveling wave (Fig. \ref{fig:raster}(b))\cite{York,Kilpatrick2009_PhysicaD}.

In short, we found six states of activity in the ring network with synaptic depression: paramagnetic, ferromagnetic, bump, oscillatory uniform, rotating bump and oscillatory bump. 

\subsubsection{Hopf and Turing instability of inhomogeneous solution}
\label{subsubsec:MH Results of the stab analysis}
We considered the stability of the inhomogeneous steady state solution (the B state) in order to identify the destabilization mechanisms leading to the two inhomogeneous oscillatory states, the RB and OB states. 
To analyze the stability of a network with $N=10^3$, we computed the eigenvalues of the Jacobian matrix $H$ (eq. (\ref{H}), $2000\times2000$) for frequency perturbations $\delta \hat{m}_k$ and $\delta \hat{X}_k$ ($-5\times10^2\le k\le5\times10^2-1$). Note that the highest Fourier components are $\hat{m}_{\pm5\times10^2}$ and $\hat{X}_{\pm5\times10^2}$ in a network with $N=10^3$, as determined by the sampling theorem. 

To begin our analysis, we consider a network with a fixed degree of synaptic depression ($\gamma=1.5$) and a time constant ($\tau=3$), in which the RB state occurs near the transition point between the P and B states. Fig. \ref{fig:MD_stab_J1var_ga15_tau3}(a) shows how the amplitude of localized activity, $\hat{m}_1$, depends on $\beta J_1$ for $\beta J_0=0$. The solid and dashed lines represent stable and unstable solutions to eq. (\ref{MD_selfeq_macro}). Fig. \ref{fig:MD_stab_J1var_ga15_tau3}(b) shows the distribution of eigenvalues for Jacobian matrix $H$. 
There are two mechanisms of destabilization that lead to the RB state. 
First, as $\beta J_0$ crosses $4.5$ from below, the P state becomes unstable and a Turing-Hopf instability leads to the RB state occurring because $\mid \lambda_0^{\pm}\mid<1$, $\mid \lambda_1^{\pm}\mid>1$ and $\mathrm{Im}(\lambda_1^{\pm})\neq0$. 

Next, we show that as $\beta J_1$ crosses $8$ from above, the B state becomes unstable and a Turing instability leads to the RB state. 
Fig. \ref{fig:MD_stab_J1var_ga15_tau3}(b) shows that there are eigenvalues continuously distributed between $\lambda=0.15$ and $0.65$ that do not contribute to destabilization and that there are a few eigenvalues greater than $1$ that do. 
 Figs. \ref{fig:MD_stab_J1var_ga15_tau3}(c) and (d) show the eigenvector of the largest eigenvalue for $\beta J_0=0$ and $\beta J_1=6.5$. It is indicated by the $\times$ mark in Fig. \ref{fig:MD_stab_J1var_ga15_tau3}(b). We see from Figs. \ref{fig:MD_stab_J1var_ga15_tau3}(c) and (d) that their eigenvectors mainly consist of $\delta\hat{m}_1$, $\delta\hat{X}_1$, $\delta\hat{m}_{-1}$, and $\delta\hat{X}_{-1}$ and not $\delta\hat{m}_0$ or $\delta\hat{X}_0$. In addition, they do not have an imaginary part. We hence found that the B state was unstable and that the RB state occurred due to a Turing instability.

\begin{figure}[!h]	
\centering
\includegraphics[width=4.5in,clip]{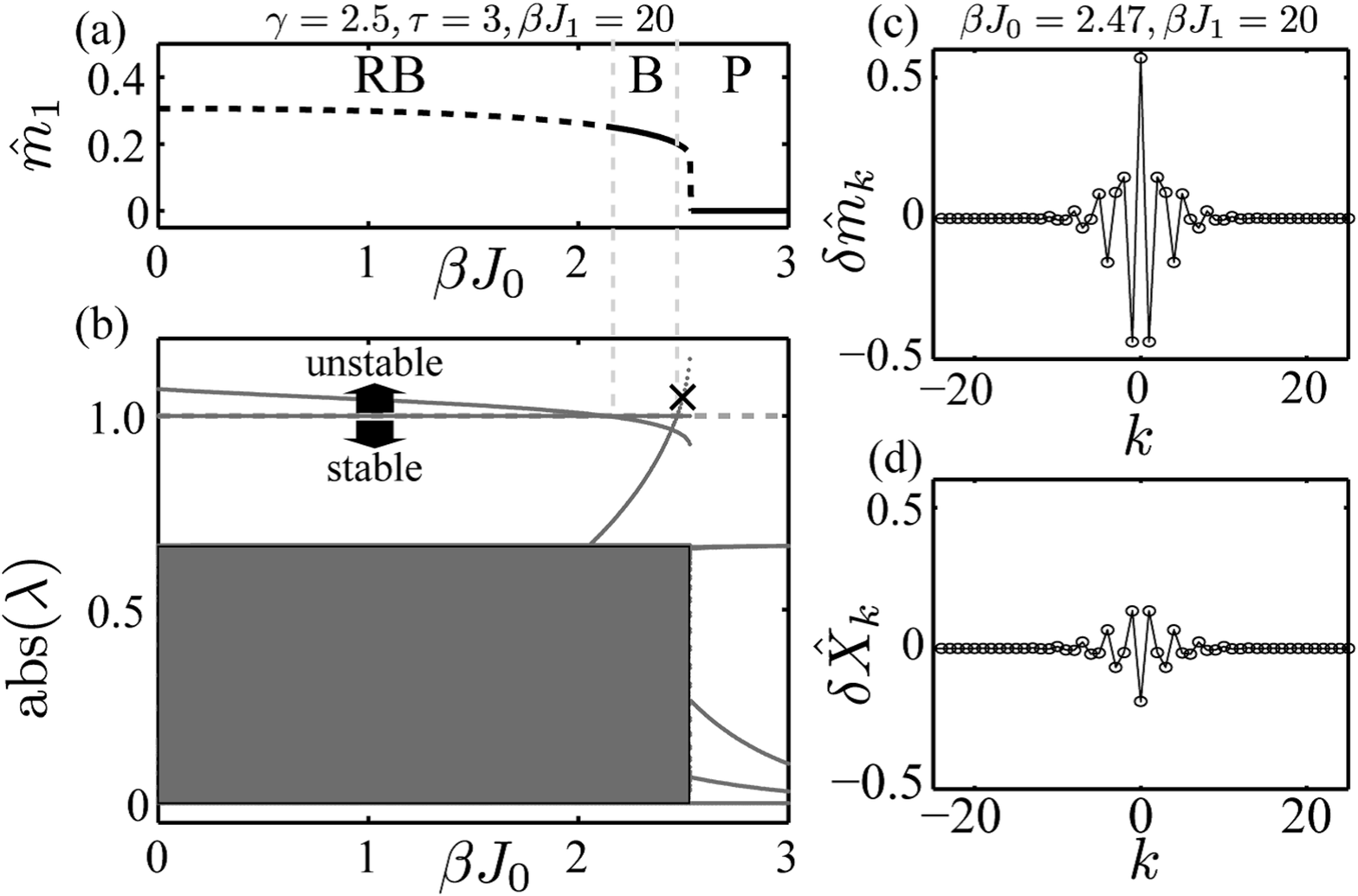}
\caption{(a) Stable (solid lines) and unstable solutions (dashed line) to amplitude of localized activity $\hat{m}_1$. (b) Distribution of eigenvalues for Jacobian matrix $H$(\ref{H}) with size of $2000\times2000$. (c), (d) Eigenvector with largest eigenvalue, $1.04$ for $\beta J_0=2.47$ and $\beta J_1=20$, as shown $\times$ in (b).}
\label{fig:MD_stab_J0var_ga25_tau3}
\end{figure}

Next, we consider a ring network with $\gamma=2.5$ and $\tau=3$, in which the B state is unstable and the OB state occurs at the transition point ($\beta J_0=2.47$) between the B and P states. As $\beta J_0$ crosses $2.47$ from below, the OB state becomes unstable and the P state occurs. 
Fig. \ref{fig:MD_stab_J0var_ga25_tau3}(a) shows how the amplitude of localized activity $\hat{m}_1$ depends on $\beta J_0$ for $\beta J_1=20$. 
Fig. \ref{fig:MD_stab_J0var_ga25_tau3}(b) shows the distribution of eigenvalues for Jacobian matrix $H$. Only a few eigenvalues are greater than $1$, as they were in Fig. \ref{fig:MD_stab_J1var_ga15_tau3}(b). Figs. \ref{fig:MD_stab_J0var_ga25_tau3}(c) and (d) show the eigenvectors of the largest eigenvalue for $\beta J_0=2.47$ and $\beta J_1=20$ indicated by the $\times$ mark in Fig. \ref{fig:MD_stab_J0var_ga25_tau3}(b), which indicates that the eigenvectors mainly consisted of $\delta\hat{m}_0$ and $\delta\hat{X}_0$. Furthermore, they had an imaginary part. The B state was thus unstable, and a Hopf bifurcation led to the OB state for $\beta J_0=2.47$.

These results clearly show that there were two mechanisms destabilizing the B state, i.e., a Turing instability and a Hopf instability, which led to the RB and the OB states. Moreover, we found that there were few eigenvalues that were larger than $1$ and thus affected the stability of the system, as shown in Figs. \ref{fig:MD_stab_J1var_ga15_tau3} and \ref{fig:MD_stab_J0var_ga25_tau3}, and that their eigenvectors consisted of only low-frequency Fourier components.

\subsubsection{Phase diagram}
\begin{figure}[!h]
\centering
\includegraphics[width=4.5in]{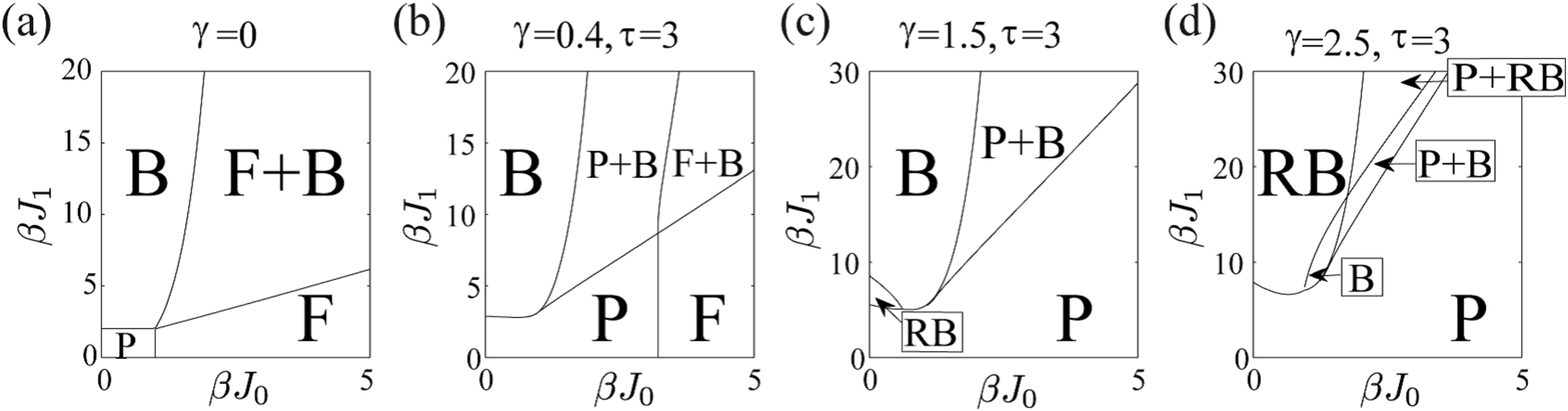}
\caption{Phase diagrams with fixed $\gamma$ in interaction of ($\beta J_0$, $\beta J_1$) plane for $\gamma=\{0, 0.4, 1.5, 2.5\}$. For (b) to (d), $\tau=3$. 
P : monostability in $\hat{m}_0$, and $\hat{m}_1=0$. 
F: bistability in $\hat{m}_0$, and $\hat{m}_1=0$. 
B: $\hat{m}_1\neq0$. 
RB : $\hat{m}_1\neq0$. 
OB state occurs on transition line between P+B state and P state with $\gamma=2.5$ and $\tau=3$ in (d). }
\label{fig:Phase_diag}
\end{figure}

We investigated how neuron interactions affect the macroscopic states of networks by changing the strength of the uniform connections ($J_0$) and the lateral-inhibitory connections ($J_1$). Fig. \ref{fig:Phase_diag} shows the phase diagrams identified in the ($\beta J_0$, $\beta J_1$) plane with a fixed degree of synaptic depression ($\gamma=\{0, 0.4, 1.5, 2.5\}$). To analyze the stability of the steady state solutions in a network with $N=10^3$, we computed the eigenvalues of Jacobian matrix $H$ for only low-frequency perturbations, namely $\delta \hat{m}_k$ and $\delta \hat{X}_k$ ($-5\times10\le k\le5\times10-1$), since high-frequency perturbations do not affect the stability of the system when there is weak lateral-inhibitory interaction ($J_1$) (See Appendix). The instability of an oscillatory state was numerically investigated using eqs. (\ref{meanf_m}) and (\ref{meanf_X}) with $N=10^3$.

To begin with, we describe the behavior of a network with nondepressed synapses ($\gamma=0$). Fig. \ref{fig:Phase_diag}(a) shows the phase diagram for $\gamma=0$\cite{Hamaguchi2006}. The relative strength of $\beta J_0$ and $\beta J_1$ determines the network state. The F or B states become stable once $\beta J_0$ and $\beta J_1$ exceed certain thresholds. Between these two states, there are bistable regions where both F and B states are locally stable (F+B). If both $\beta J_0$ and $\beta J_1$ are small, a P state is stable.

In the presence of weak synaptic depression ($\gamma=0.4$, $\tau=3$) (Fig. \ref{fig:Phase_diag}(b)), the P region expanded and the bistable regions shrunk since synaptic depression effectively reduced the lateral-inhibitory ($J_1$) interaction. Bistable regions, where both P and B states were locally stable (P+B), developed. As the degree of synaptic depression increased ($\gamma=1.5$, $\tau=3$), the RB state became stable near the transition point between the P and B states, as shown in Fig. \ref{fig:Phase_diag}(c). Since $\gamma>1$, the F state was unstable.

In the presence of strong synaptic depression ($\gamma=2.5$, $\tau=3$) (Fig. \ref{fig:Phase_diag}(d)), the RB region expanded and the P+B region shrunk. Bistable regions, where both the P and RB states are locally stable (P+RB), developed. The B state was unstable, and an OB region developed at the transition between the B+P and P states (Fig. \ref{fig:MD_mj_Xj_Jij}(g)$\sim$(i)). Otherwise, the OB state was unstable, and the P state was stable.

We have shown that for a sufficiently strong degree of synaptic depression, the B state is unstable and the oscillatory states (RB and OB) occur near the transition of the B and P states. Sufficiently strong lateral inhibition interaction ($J_1$) leads to a Turing instability and the RB state occurs. Sufficiently strong uniform inhibition interaction $J_0$ leads to a Hopf instability, leading to the OB state.

\subsection{Summary}
In the ring attractor network with synaptic depression, there are homogeneous steady states (F, P) and an inhomogeneous steady state (B). We have shown that, depending on the strength of the interneuron connections, instability in these states leads to three oscillatory states: oscillatory uniform (OU), rotating bump (RB), and oscillatory bump (OB).

\begin{figure}[!h]
\centering
\includegraphics[width=4.5in,clip]{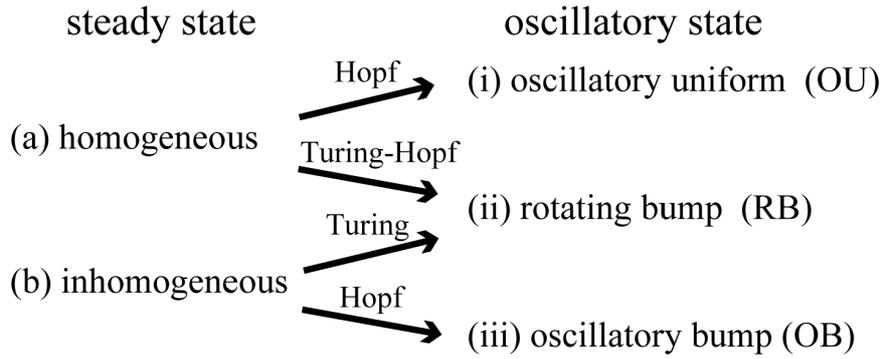}
\caption{Schematic view of stability analysis for ring network with synaptic depression.}
\label{summary_of_StabAnalysis}
\end{figure}

We summarize how the interactions between neurons affect the stabilities of the steady state solutions by changing the strengths of the uniform connections ($J_0$) and the lateral-inhibitory connections ($J_1$). Fig. \ref{summary_of_StabAnalysis} shows a schematic view of the stability analysis for a ring network with synaptic depression.
First, for sufficiently weak uniform connections and a sufficiently long time constant ($\tau$), the homogeneous steady state solution, which has a high firing rate was unstable near the transition point between the P and F states. An OU state then developed due to a Hopf bifurcation, as explained in \S\ref{subsubsec:Oscillatory states} (Figs. \ref{summary_of_StabAnalysis}(a)$\to$(i)). 
Note that the OU state developed in the presence of lateral-inhibitory connections as well as in their absence. 
Next, as the strength of the uniform connections increased, the inhomogeneous steady state solution became unstable near the transition point between the B and P states (Fig. \ref{fig:MD_stab_J0var_ga25_tau3}). The OB state then developed because of a Hopf instability, as explained in \S\ref{subsubsec:MH Results of the stab analysis} (Figs. \ref{summary_of_StabAnalysis}(b)$\to$(iii)). In the OB state, bump state activity moved up and down around the most activated neuron (Figs. \ref{fig:MD_mj_Xj_Jij}(g)-(i)). 
Finally, homogeneous state P and inhomogeneous state B became unstable near the transition point between them (Fig. \ref{fig:MD_stab_J1var_ga15_tau3}(a)), and the RB state developed, as explained in \S\ref{subsubsec:MH Results of the stab analysis} (Figs. \ref{fig:MD_mj_Xj_Jij}(d)-(f)). As the strength of the lateral-inhibitory connections increased, the homogeneous steady state solution became unstable, and a RB state developed because of a Turing-Hopf instability (Figs. \ref{summary_of_StabAnalysis}(a)$\to$(ii)). As the strength of the lateral-inhibitory connections decreased, the inhomogeneous steady state solution became unstable and an RB state developed due to a Turing instability (Figs. \ref{summary_of_StabAnalysis}(b)$\to$(ii)). 

These results show that medium-strength uniform connections cause a Hopf instability near the transition point and up-and-down movement of the firing rates and the average synaptic variables, and that medium-strength lateral-inhibitory connections cause a Turing instability near the transition point and propagation of a localized bump of activity around the ring network. In sum, various oscillatory states take place depending on the strength of the interneuron connections in a ring network with synaptic depression.

\section{Conclusion}
\label{sec:Conclusion}
We have explored the macroscopic properties of two types of stochastic binary neural networks with synaptic depression: a network with homogeneous connectivity and a ring attractor network with Mexican-hat type connectivity. We proposed a dynamical mean field theory for a stochastic binary neural network model with synaptic depression assuming that synaptic weight $J_{ij}$ is of the order of $1/N$ with respect to the number of neurons ($N$). Using microscopic mean field equations, we derived macroscopic steady state equations for these networks and investigated the stability of the steady state solutions obtained. The results coincided with those from simulation. 
We conclude that the presence of synaptic depression leads to oscillatory instability and that various oscillatory states take place depending on the strength of the interneuron connections. Synaptic depression thus causes a diversity of dynamic states in large networks of spiking neuron.

We focused only on non-frustrated systems, in which $J_{ij}\sim O(1/N)$. A further direction of this study will be to extend the microscopic dynamical mean field theory to frustrated systems, such as the Sherrington and Kirkpatrick model\cite{SKmodel}.

\section{Appendix: Dimensionality reduction for stability analysis}
\label{sec:Dimensionality reduction}
\begin{figure}[!h]	
\centering
\includegraphics[width=4.5in,clip]{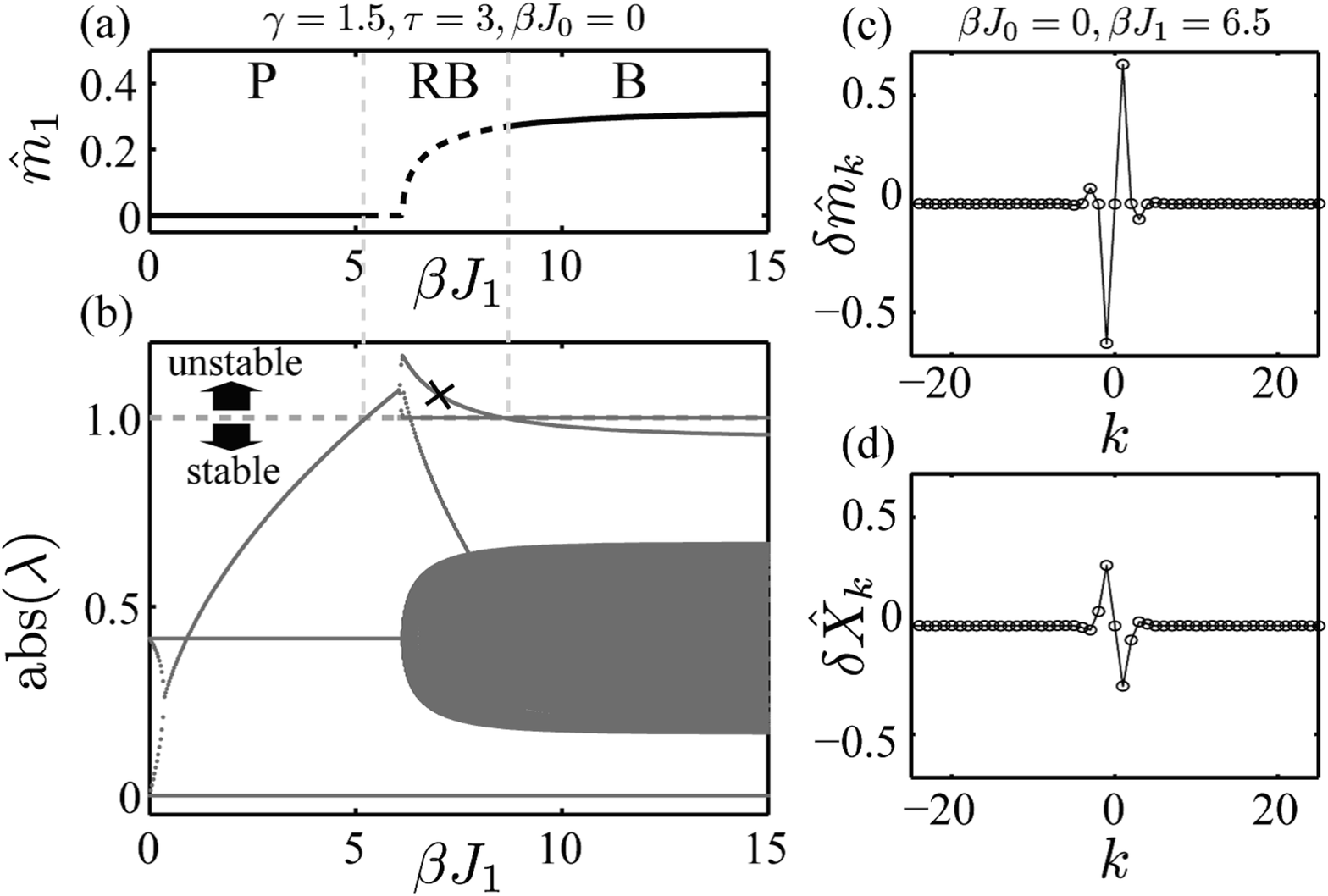}
\caption{(a) Stable (solid lines) and unstable solutions (dashed line) to amplitude of localized activity $\hat{m}_1$. (b) Distribution of eigenvalues for Jacobian matrix $H$(\ref{H}) with size of $2000\times2000$. (c), (d) Eigenvector with largest eigenvalue, $1.1$ for $\beta J_0=0$ and $\beta J_1=6.5$, as shown by $\times$ markin (b).}
\label{fig:MD_stab_J1var_ga15_tau3}
\end{figure}

\begin{figure}[!h]
\centering
\includegraphics[width=4.5in,clip]{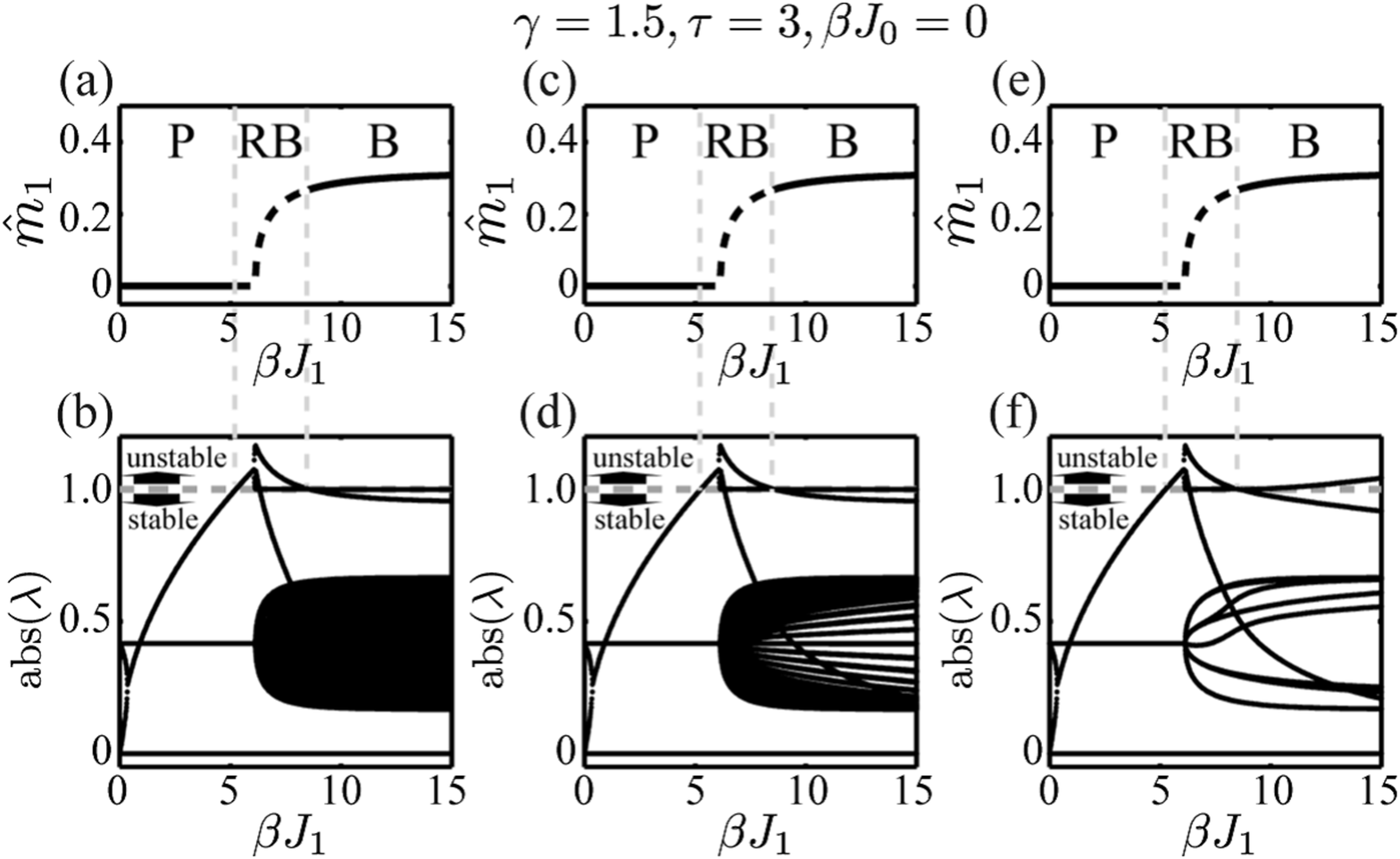}
\caption{Solid and dashed lines in (a), (c), and (e) represent stable and unstable solutions to the amplitude of localized activity, $\hat{m}_1$. (b), (d), (e) Distribution of eigenvalues for Jacobian matrix $H$ (eq. \ref{H}) for perturbations $\delta\hat{m}_k$ and $\delta\hat{X}_k$ ($-K\le k\le K-1$). (b) $K=5\times10^2$. The size of the Jacobian matrix $H$ is $2000\times2000$. (d) $K=5\times10$. The size of the Jacobian matrix $H$ is $200\times200$. (e) $K=5$. The size of the Jacobian matrix $H$ is $20\times20$.}
\label{fig:compare_approx_J0const}
\end{figure}

\begin{figure}[!h]
\centering
\includegraphics[width=4.5in,clip]{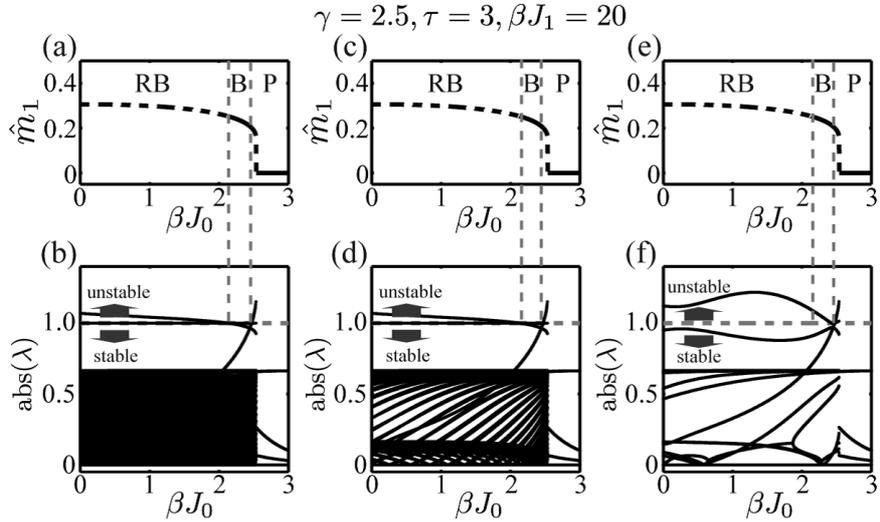}
\caption{Solid and dashed lines in (a), (c), (e) represent stable and unstable solutions to the amplitude of localized activity $\hat{m}_1$. (b), (d), (e) Distribution of eigenvalues for the Jacobian matrix $H$ (eq. \ref{H}) for perturbations $\delta\hat{m}_k$, $\delta\hat{X}_k$ ($-K\le k\le K-1$). (b) $K=5\times10^2$. Size of Jacobian matrix $H$ is $2000\times2000$. (d) $K=5\times10$. Size of Jacobian matrix $H$ is $200\times200$. (e) $K=5$. Size of Jacobian matrix $H$ is $20\times20$.}
\label{fig:compare_approx_J1const}
\end{figure}

In Section \ref{subsubsec:MH Results of the stab analysis}, we showed that few eigenvalues consisting of low-frequency perturbations affected the stability of the system. This means that only low-frequency perturbations affect the stability of the system. In this section, we compare the eigenvalues of the Jacobian matrix in Fourier space (eq. (\ref{H})) for high and low frequency perturbations with those for low-frequency perturbations, as indicated in Figs \ref{fig:compare_approx_J0const} and \ref{fig:compare_approx_J1const}. Fig. \ref{fig:compare_approx_J0const}(b) shows the distributions of eigenvalues for the Jacobian matrix (eq. \ref{H}) for perturbations $\delta\hat{m}_k$ and $\delta\hat{X}_k$ ($-5\times10^2\le k\le5\times10^2-1$) in a ring network with $\gamma=1.5$ and $\tau=3$ as well as Fig. \ref{fig:MD_stab_J1var_ga15_tau3}(b). Figs. \ref{fig:compare_approx_J0const}(d) and (e) show those for low-frequency perturbations, namely $\delta\hat{m}_k$ and $\delta\hat{X}_k$ ($-5\times10\le k\le5\times10-1$), and $\delta\hat{m}_k$ and $\delta\hat{X}_k$ ($-5\le k\le5-1$) respectively. Fig. \ref{fig:compare_approx_J1const} shows a comparison for a ring network with $\gamma=2.5$ and $\tau=3$ in the same way.

Although we did not approximate the distributions in Fig. \ref{fig:compare_approx_J0const} (b) as a whole by using the distributions shown in Figs. \ref{fig:compare_approx_J0const} (d), we found that the maximum eigenvalue of the Jacobian matrix for perturbations $\delta \hat{m}_k$ and $\delta \hat{X}_k$ ($-5\times10\le k\le5\times10-1$) coincided with that of the Jacobian matrix for perturbations $\delta \hat{m}_k$ and $\delta \hat{X}_k$ ($-5\times10^2\le k\le5\times10^2-1$). However, the results of stability analysis for perturbations $\delta \hat{m}_k$ and $\delta \hat{X}_k$ ($-5\le k\le5-1$) and those for perturbations $\delta \hat{m}_k$ and $\delta \hat{X}_k$ ($-5\times10^2\le k\le5\times10^2-1$) differed in the maximum eigenvalue of the Jacobian matrix because the lateral-inhibitory interaction ($J_1$) was relatively strong (Figs. \ref{fig:compare_approx_J0const} (b), and(f)). In contrast to the stability analysis for $\delta \hat{m}_k$ and $\delta \hat{X}_k$ ($-5\times10\le k\le5\times10-1$), we obtained different results for the stability analysis. We obtained the same results of stability analysis, as shown in Fig. \ref{fig:compare_approx_J1const}.

These results show that high-frequency perturbations $m_k$ and $X_k$ did not affect the stability of inhomogeneous steady state solutions because the lateral-inhibitory interaction ($J_1$) was relatively weak. Therefore, when there is weak lateral-inhibitory interaction, we can reduce the dimensions for the stability analysis since we do not need to take into account high-frequency perturbations. This dimensionality reduction enables rapid analysis of the stability of steady state solutions for a ring network with synaptic depression.

\clearpage


\end{document}